\documentclass[useAMS,usenatbib]{mn2e} 
\usepackage{graphicx}
\usepackage{natbib}
\usepackage{times}
\usepackage{xcolor}

\def\cM{{\cal M}}
\def\Teff{T_{\rm eff}}
\def\logg{\log g}
\def\feh{\hbox{[Fe/H]}}
\def\mh{\hbox{[M/H]}}

\def\ex#1{\left\langle#1\right\rangle}
\newcommand{\deltanu}{\mbox{$\Delta\nu$}}
\newcommand{\numax}{\mbox{$\nu_{\rm max}$}}

\title[Distance and extinction to APOGEE stars]
{Distance and extinction determination for APOGEE stars with Bayesian method}
\author[Jianling Wang et al.]{
Jianling Wang$^{1,2}$\thanks{E-mail: wjianl@bao.ac.cn},
Jianrong Shi$^{1,2}$,
Kaike Pan$^{3}$,
Bingqiu Chen$^{4}$,
Yongheng Zhao$^{1,2}$,\newauthor
James Wicker$^{2}$ \\
$^{1}$ Key Laboratory of Optical Astronomy, NAOC, Chinese Academy of Sciences\\
$^{2}$ National Astronomical Observatories, Chinese Academy of Sciences, Beijing 100012, China\\
$^{3}$ Apache Point Observatory and New Mexico State University, P.O. Box 59, Sunspot, NM, 88349-0059, USA \\
$^{4}$ Department of Astronomy, Peking University, Beijing 100871.\\
}

\begin{document}
\date{Accepted ... Received ...}
\pagerange{\pageref{firstpage}--\pageref{lastpage}} \pubyear{2008}
\maketitle
\label{firstpage}

\begin{abstract}

Using a Bayesian technology we derived distances and extinctions for over
$100,000$ red giant stars observed by the Apache Point Observatory Galactic
Evolution Experiment (APOGEE) survey by taking into account spectroscopic
constraints from the APOGEE stellar parameters and photometric
constraints from 2MASS, as well as a prior knowledge on the Milky Way. Derived
distances are compared with those from four other independent methods, the {\it
Hipparcos} parallaxes, star clusters, APOGEE red clump stars, and asteroseismic
distances from APOKASC \citep{rodrigues14} and  SAGA Catalogues
\citep{casagrande14}.  These comparisons covers four orders of magnitude in the
distance scale  from 0.02 kpc to 20 kpc.  The results show that our distances
agree very well with those from other methods: the mean relative difference
between our Bayesian distances and those derived from other methods ranges from
$-4.2\%$ to $+3.6\%$, and the dispersion ranges from 15\% to 25\%. The
extinctions toward all stars are also derived and compared with those from
several other independent methods: the Rayleigh$-$Jeans Color Excess
(RJCE) method, Gonzalez's two-dimensional extinction map, as well as
three-dimensional extinction maps and models. The comparisons reveal that,
overall, estimated extinctions agree very well, but RJCE tends to overestimate
extinctions for cool stars and objects with low $\logg$.

\end{abstract}


\section{Introduction}

Galactic formation and evolution is one of the most outstanding problems in
modern astrophysics. One way to address this question is through Galactic
Archaeology, which is an approach to explore the formation and evolution
history of the Milky Way through the $``$archeological'' record provided by its
individual stars. We are coming into a new era of Galactic investigation, in
which Galactic archaeology can be studied with massive spectroscopic surveys.
The spectra generated by these surveys are range from $10^4$ to $10^7$, for
examples, RAVE \citep{steinmetz06}, SEGUE \citep{yanny09}, APOGEE
\citep{eisenstein11,majewski10}, GALAH \citep{desilva15, freeman12}, LAMOST
\citep{zhao06,cui12}, {\it Gaia} \citep{perryman01, lindegren10}, ARGOS
\citep{ness12a, ness12b}, 4MOST \citep{dejong12}, and WEAVE \citep{dalton12}.
Valuable information, that can be extracted from these spectroscopic surveys,
include fundamental stellar parameters, chemical compositions, and radial
velocities of individual stars. With the compilations of these massive surveys
in the next decade, our knowledge of Galactic formation and evolution will be
significantly improved.

As a unique campaign among these surveys, APOGEE operates in the near-infrared
band and is able to target stars at very low Galactic latitude
\citep{zasowski13}, this penetrating the veil of interstellar dust. Besides
observing disk stars, APOGEE also targets two other components: the halo and
the bulge. It has pencil-beam observations of red giants in the halo that
target fields in the bulge. The primary goal of APOGEE is to provide
constraints on the dynamical and chemical evolution model of our Milky Way.
Numerous Galactic structure and stellar population issues can be addressed by
the APOGEE data.  In order to take full advantage of this massive spectroscopic
data set, and make a full use of the chemical information, we need to analyze
the 6D phase-space distribution of stars in the Milky Way, which is usually
hampered by a lack of reliable distance data.

The most accurate and successful direct distances measurements are from the
{\it Hipparcos} satellite, which provide trigonometric parallaxes for $\sim
10^5$ stars. However, the {\it Hipparcos} measurements are only accurate to a
distance around $200$ pc \citep{burnett10}. In the near future, the recently
launched {\it Gaia} mission (Perryman 2005) will return parallax and proper
motion measurements for around $10^9$ stars, which will contribute to this
field substantially. However, even after the {\it Gaia} mission is completed,
stars with direct distance measurements are still less than 1\% of all stars in
the Milky Way. Indirect distance determination methods for more distant stars,
especially for APOGEE stars with heavy extinction in the low latitude disk
region, and in the bulge, are still highly desired.

Several indirect methods for determining stellar distance have been developed.
Red Clump (RC)  stars has been used as standard candles to calculate stellar
distances \citep{bovy14, paczynski98}.  Since the RC stage is very short in the
whole stellar evolutionary life of a star, in general, this method is only
suitable for a small fraction of stars. Asteroseismology can determine stellar
parameters, such as stellar mass, radius, and age \citep{casagrande14}, with
high precision. These parameters are then employed, together with other stellar
parameters determined with classical methods, e.g., metallicity and effective
temperature, to accurately estimate intrinsic stellar properties, such as
distance and extinction. \citet{rodrigues14} have clearly demonstrated that
distance can be obtained with high accuracy when coupled with asteroseismic
information. However, the Asteroseismology method requires systematic
variability studies of its targets. It not only needs the target stars to have
detectable seismic activities, but also requires observing a very time
consuming data set. Till now, only small patch of sky coverage has become
available with Asteroseismology data \citep{koch2010}, and many stars within
the coverage area do have asteroseismic data for accurate distance
determinations. Therefore, more general methods of deriving stellar distance
are needed.

To take advantage of large sets of available spectroscopic and photometric data
from recent surveys, more general methods to infer stellar parameters, such as
extinction, mass, age, distance, etc, have been explored by many authors
\citep{pont04, schonrich14, wang2015}. Because of its importance, many among
these works focus on distance determination \citep{breddels10, zwitter10,
burnett10, binney14, santiago15, wang2015}. These authors use spectroscopic and
photometric quantities to compute the probability distribution of stellar
parameters, to infer distance of individual stars. The logistics of these works
are similar, but different authors may implement the method in slightly
different ways. For instance, some use estimated uncertainties of measured
stellar parameters, some do not. Authors may treat extinction differently, use
different theoretical evolutionary models, use different prior knowledge of the
Galaxy, etc.

In this work, we follow the same Bayesian method as \citet{wang2015} to derive
distances for around $10^5$ stars in the APOGEE survey.  The extinction is
considered consistently along with the distance determination. The plan for the
paper is as follows. In Section 2, we briefly introduce the Bayesian method
adopted in this work, while the APOGEE data from SDSS DR12 are described in
Section 3. In Sections 4 and 5, we compare our Bayesian distances and
extinctions with those from other independent methods.  Finally, the conclusion
is presented in Section 6.

\section{Bayesian method}
\label{sec:method}

The Bayesian method adopted here is the same as used by \citet{wang2015}, in
which the distances and extinctions were calculated for around one million
stars in the first data release of the LAMOST survey \citep{luo15}. The
accuracy of the derived distances have been extensively tested with both
simulation and several independent measurements utilizing LAMOST data. The
check shows that a good accuracy can be achieved with the implemented Bayesian
method in \citet{wang2015}. The method is very similar to those in
\citet{burnett10} and \citet{binney14}. For clarity we give a brief description
on this method.

For each star we have the relevant observables: effective temperature $\Teff$,
metallicity $\mh$, surface gravity $\logg$, and near infrared
photometries $J, H, K_s$. These quantities form an observed vector:
\begin{equation}
\mathbf{O}=(\Teff, \mh, {\log g}, J, H, Ks)
\end{equation}
Each star can be related and characterized by a set of
$``$intrinsic'' parameters: metallicity $\mh$, age $\tau$, initial
mass $\cM$, position on the sky $l, b$, and distance from the Sun
$d$. These quantities form another vector:
\begin{equation}
 \mathbf{X}=(\mh, \log \tau, \cM, l, b, d)
\end{equation}
With the help of trivial Bayesian theory, we can derive the posterior
probability of $\mathbf{P(X|O})$, which is the conditional probability of the
parameter set $\mathbf{X}$  given $\mathbf{O}$.

\begin{equation}
 P({\mathbf{X|O}}) = \frac{P({\mathbf{X}})}{P({\mathbf{O}})}P({\mathbf{O}}|{\mathbf{X}}),
\end{equation}

$\mathbf{O}$ and $\mathbf{X}$ can be connected by theoretical isochrones, with
$\mathbf{O}$ being a function of $\mathbf{X}$.  $\mathbf{P(O)}$ is the
probability that the set of observations was made and does not depend on
$\mathbf{X}$, which is a normalization factor.  $\mathbf{P(O|X)}$ is the
probability that the set of observations $\mathbf{O}$ gives the set of
parameters $\mathbf{X}$.  We assume the uncertainties of the measured
observables for each component of $\mathbf{O}$ can be described as a Gaussian
distribution with a mean $\mathbf{\tilde{O}}$ and standard deviation
$\mathbf{\sigma_{\tilde{O}}}$.

\begin{equation}
P(\mathbf{\tilde{O}|X}) = G(\mathbf{\tilde{O}|O(X),\sigma_o})= \prod_{(i=1)}^n G(\tilde{O_i}|\mathbf{O(X)},\sigma_{oi})
\end{equation}

$\mathbf{P(X)}$ is the prior probability we ascribe to the set of parameters,
which is an important ingredient in the Bayesian method.  \citet{burnett10}
used a three-component prior model of the Galaxy for the distribution functions
of metallicity, density, and age:

\begin{equation}
P(\mathbf{X}) = p({\cal M}) {\sum_{i=1}^3} p_i(\mh)  p_i(\tau) p_i(\mathbf{r}) A_{V{\rm prior}}(\ell,b,d)
\end{equation}

Where the $i=1,2,3$ correspond to thin-, thick-disk, and stellar halo,
respectively. A slightly modified Kroupa-type IMF is adopted as shown in
\citet{burnett10} and \citet{binney14}. The same as \citet{binney14},  an
extinction prior, which employs the three-dimensional Milky Way extinction
model, is used to calculate the extinction by integrating along each line of
sight toward individual stars.

Having the posterior probability distribution function,
$P({\mathbf{X|\tilde{O}}})$, the mean and standard deviation for each
parameters in $\mathbf{X}$ can be obtained by taking the first and second
moments of this distribution as shown by \citet{burnett11}. Besides providing
the distance $(\ex{d})$, this Bayesian method can also output another distance
estimator, the parallax $(\ex{\varpi})$. In general, there is the relation
between these two distance estimators, $\ex{d}$ ${\ga}$ $1/\ex{\varpi}$, which
can be attributed to the weights that each estimator attaches to the
possibilities of long or short distance\citep{binney14}.  \citet{binney14} find
that $1/\ex{\varpi}$ is a good distance estimator for RAVE data.  In this work,
we will use the output $\ex{d}$ directly in the following sections. However,
for comparison we also list the results estimated from $\ex{\varpi}$ in
Table.\ref{table:d}.

The Padova isochrones\footnote{http://stev.oapd.inaf.it/cgi-bin/cmd}
\citep{marigo08, marigo07, girardi00, bertelli94} are adopted like in
\citet{wang2015}.  The metallicity steps of the isochrones have been carefully
chosen by \citet{burnett10} for avoiding great changes in the observed
quantities between adjacent isochrones, and have been extended to low
metallicity by \citet{wang2015}.

Since the extinction can be calculated with this Bayesian method, an extinction
law is a necessary ingredient. At present, there is no real consensus on the
correct extinction law along all the directions of the Galaxy, especially
toward to the Galactic bulge region
\citep{gonzalez12,nishiyama06,nishiyama08,nishiyama09}. Thus, we decide to use
the extinction law of \citet[][CCM]{ccm89} with $R_V=3.1$ as shown in
\citet{wang2015}.


\section{Data samples}
\label{sec:sample}

The Apache Point Observatory Galactic Evolution Experiment (APOGEE) as
one component of SDSS III \citep{eisenstein11}, is a near-infrared ($H$-band;
$1.51\mu$m$\sim1.70\mu$m), high-resolution (R$\sim$22,500) spectroscopic
survey. It targets primarily red giant (RG) stars across all Galactic
environments and stellar populations in the Milky Way \citep{allende08,
majewski10}, using the Sloan 2.5 m telescope\citep{gunn06} and an
innovative multi-object IR spectrograph \citep{wilson10}.  The stellar
atmospheric parameters and individual chemical abundances are derived from
combined APOGEE spectra with the APOGEE Stellar Parameters and Chemical
Abundances Pipeline \citep[ASPCAP;][]{garciaperez15}, which finds the best
matching between pre-calculated synthetic spectra to the observed ones via a
$\chi$$^2$ minimization. To validate ASPCAP outputs, stellar parameters for
members in well studied clusters and surface gravities from asteroseismic
analyses are carefully compared \citep{meszaros13}.  Calibration correlations
are applied to the original ASPCAP outputs to obtain calibrated stellar
parameters for stars \citep{holtzman15}. The empirical uncertainty of $\Teff$
is about 100 K through comparison of ASPCAP derived temperatures with
photometric temperatures, and 0.11 dex for $\logg$ via comparing calibrated
ASPCAP values with asteroseismic surface gravities for stars in the $Kepler$
field. The internal precision of the APOGEE abundances is typically $0.05-0.1$
dex, but the external accuracy of the abundances is challenging to assess, and
may only be good to 0.1-0.2 dex \citep{holtzman15}. It is worthy to note that
the applied calibration correlations are derived with members in well-studied
clusters and stars in the $Kepler$ field; their reliability outside parameter
spaces that these calibrated stars cover,  such as surface gravities of metal
poor stars, are not validated.

The final SDSS-III APOGEE public data release, DR12 \citep{alam2015}, includes
catalogs with radial velocity, stellar parameters $(\Teff, \logg, \mh,
[\alpha/{\rm M}])$, and 15 elemental abundances for over 150,000 stars, as well
as more than 500,000 spectra from which these quantities are derived.  We
retrieved the calibrated stellar parameter data from the APOGEE Parameter
Catalogs of SDSS DR12 using the CasJobs
interface\footnote{http://skyserver.sdss.org/casjobs/}. The Two Micron All Sky
Survey (2MASS) near-infrared photometric data for the same stars in $J(1.24
\mu$m), $H(1.66\mu$m), $Ks(2.16\mu$m) are also extracted from the point-source
catalogue \citep{cutri03,skrutskie06}. The following procedure is used to
select stars with reliable parameters: First, stars should not be set any
flags indicating that the observations or analysis is bad. Second, stars
without any calibrated parameters (such as dwarfs) are excluded.  We
use SDSS CasJobs to extract the data with the following command scripts:
(aspcapflag \& dbo.fApogeeAspcapFlag('STAR\_BAD')) = 0 and Teff $>$ 0. With
these criteria, we select 104816 stars from the APOGEE catalogue.  There are
3084 stars that have been observed more than once.  For stars with multiple
observations, only data with the highest signal to noise ratio are used. This
results in 101726 stars in our final sample. We note that in the
following analysis only stars with $\Teff <5200$ K and $\logg <3.8$ are used to
select stars with reliable parameters \citep{holtzman15}, and there are 99165
stars left after applying this selection. After further strict criteria are
applied, such as $\Teff>3800$ K and ASPCAP\_CHI2 $<$ 20 (92060 stars left), no
significant change is noticed in the statistical results.

\section{External Comparisons on Derived Distances}
\label{sec:exterComP}

As discussed above, there are several independent methods which can deliver 
reliable distances for their suitable targets. For instance, distances to
nearby stars from parallax measurements; to cluster members from the
main-sequence fitting, distances of red clump (RC) stars by using them as a
distance candle; distances of giants from asteroseismic information, etc. Our
Bayesian method, described in Section 2, is a more general one. In principle,
it can be used to derive stellar distances of individual stars which have
available spectroscopic and photometric quantities. To validate our derived
stellar distances, we compare our derived distances with results from other
independent methods for common stars: astrometric parallax distances obtained
by ESA's {\it Hipparcos} mission \citep{van07}; distances to well studied
clusters,  distances of RC stars, and distances from asteroseismic analyses.
Figures 1 to 6 show these comparisons which are also summarized in Table 1. In
summary, the systematic offsets, in terms of our derived distances minus
others', range from $-4.2\%$ to $3.6\%$, with dispersions from 15\% to 25\%;
considering mutual uncertainties in our derived distances and in others', our
distances are consistent with those from other methods.  We discuss these
comparisons in more detail below.

\begin{table*}
\begin{center}
\caption{Summary the relative difference between our Bayesian
distances and reference distance samples. Two distance estimators
from Bayesian methods are used for comparison.} \label{table:d}
 \begin{tabular}{l|l|l|c|c|c|c|c|c|c}
 \hline
 & & & & \multicolumn{2}{c}{distance is derived from $1/\ex{\varpi}$} &  &\multicolumn{2}{c}{$\ex{d}$ is used for distance} \\ \cline{5-6}  \cline{8-9}
Samples& N$_{\rm stars}$ &Dist. Range & &\multicolumn{2}{c}{Relative residual(\%)}&  &\multicolumn{2}{c}{Relative residual(\%)} \\ \cline{5-6}\cline{8-9}
    &           &   (kpc)  & & Mean               &   R.M.S.              &  & Mean & R.M.S.  \\
 \hline
Hipparcos       &  662 & 0.02-0.2& & $-3.8\pm1.0$& 25 & & $-0.6\pm1.0$ &25  \\
APOGEE-clusters &  431 & 0.8-19.3& & $-5.5\pm1.0$& 20 & & $-4.2\pm1.1$ &18 \\
APOGEE-RC       & 19937& 0.4-12.5& & $-3.3\pm0.1$& 17 & & $-1.2\pm0.1$ &18 \\
Rodrigues et al.& 1989 & 0.3-4.7 & & $+1.2\pm0.3$& 15 & & $+3.6\pm0.3$ &16 \\
SAGA            &  135 & 0.7-4.1 & & $-2.1\pm0.7$& 15 & & $+0.0\pm0.8$ &15 \\

\hline
\end{tabular}
\end{center}
\end{table*}

\subsection{Samples with Accurate {\it Hipparcos} Distances}
\label{sec:resHip}

\begin{figure*}
\centering
\includegraphics[width=8.1cm]{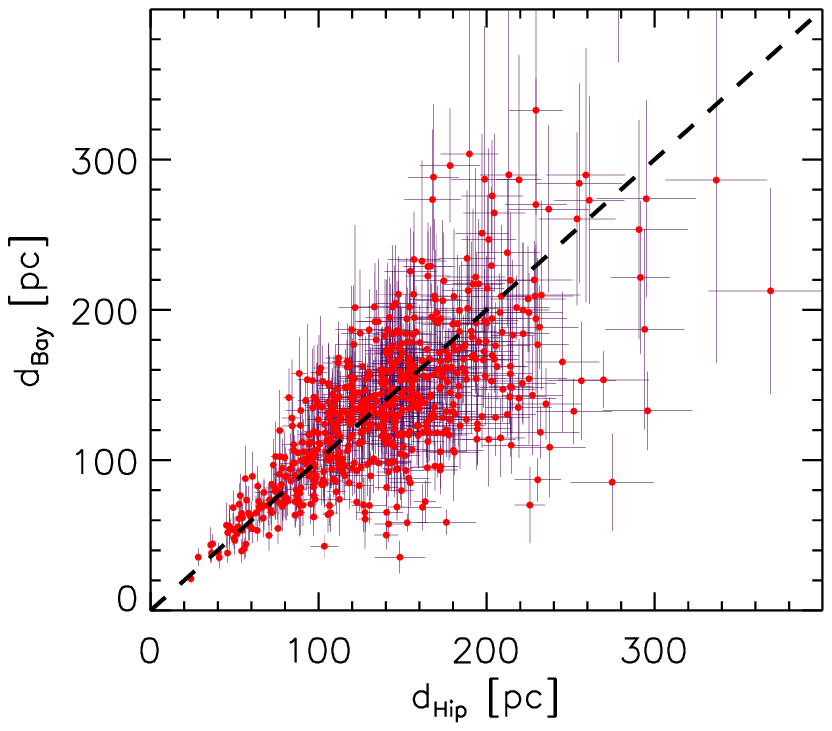}
\includegraphics[width=8.1cm]{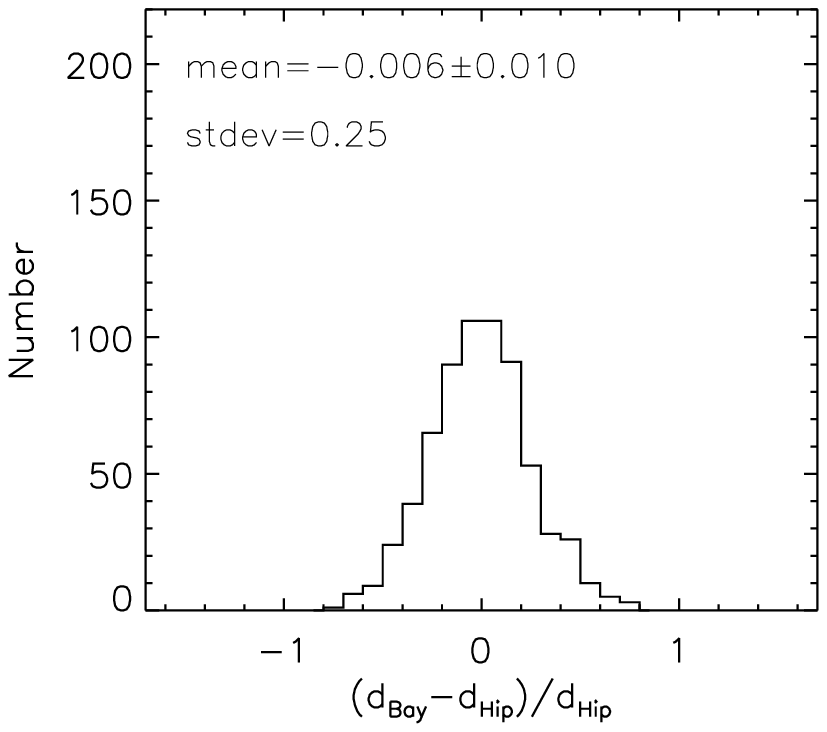}
\caption{ 
Left panel: comparison of distances to stars derived from our Bayesian method
(d$_{\rm Bay}$) with those from the {\it Hipparrocs} parallaxes (d$_{\rm Hip}$)
with $\sigma_{\varpi}/\varpi<0.1$. The horizontal error bars are
derived from uncertainties of the {\it Hipparrocs} parallaxes, while the
vertical error bars are from the formal error outputted by the Bayesian method
explained in Section 2. Right panel: fractional difference distribution of
distances with mean value and dispersion labeled on the top. The error of the
mean value in the right panel is estimated by the bootstrap method with 1000
samples. }
\label{fig:hip}
\end{figure*}

The {\it Hipparcos} mission measured trigonometric parallaxes of $\sim 10^5$
nearby stars. \citet{van07} improved the measured distances of the {\it
Hipparcos} stars through a re-reduction process. It is believed that the
\citet{van07} catalogue contains the most accurate set of distance measurements
for nearby stars ($d < 300$ pc). Therefore, here we compare our derived
distances with those from the \citet{van07} catalogue for common stars between
APOGEE data set and {\it Hipparcos} stars.

An ancillary APOGEE project, observing bright nearby stars which have {\it
Hipparcos} parallax measurements,  is setup. It would be a very inefficient way
to use available resources if the regular APOGEE survey mode, the Sloan 2.5m
telescope + APOGEE spectrograph, is used to observe these bright objects
because there are not enough bright objects for all 300 fibers in a field and
this requires a high observational overhead due to short exposure time.
Therefore, 10 fibers were installed to connect the APOGEE instrument to the
NMSU (New Mexico State University) 1m telescope.  This configuration allows one
science fiber and nine sky fiber per observation.  Bright stars with magnitudes
of $0 < H < 8$ are observed in this configuration. The NMSU 1m$+$APOGEE is used
during dark time when the APOGEE instrument is not used with the Sloan 2.5m
Telescope (see Feuillet et al, 2015 for more details).

The spectra taken with the NMSU 1m$+$APOGEE are reduced and analyzed with the
ASPCAP pipeline \citep{holtzman15}. Compared to the main survey data,
a different method was needed to treat the telluric absorption because no hot,
relatively featureless star could be observed simultaneously. The atmospheric
model spectrum is combined with a spectral template that best fits the target
and is adjusted to fit the telluric features in the observed target spectrum.
This process is iterated to produce the telluric absorption spectrum that best
matches the observed spectrum \citep{feuillet15}.  Similar to the main survey,
sky emission features are subtracted with data from sky fibers. Because there
are more sky fibers which can also be placed closer to the target star, in the
NMSU 1m$+$APOGEE observations, presumably, sky emission features can be removed
better. The stellar parameters and abundances are then derived from the spectra
with ASPCAP \citep{garciaperez15}. Overall, stellar parameters and
abundances for stars that are observed with the NMSU 1m$+$APOGEE are obtained
in an almost identical way as for the APOGEE main survey.

Hundreds of stars with parallax error $< 10\%$ were selected from the {\it
Hipparcos} Catelog \citet{van07}. 750 of them were observed with the NMSU
1m$+$APOGEE and included in SDSS DR12. However, 45 of them that are outside the
ASPCAP calibration range (${\log g} > 3.8, (J-K) > 0.5)$ are excluded from
\citet{feuillet15}'s analysis. We thank Feuillet for sharing this target list.
For these 705 stars, the mean difference between our derived distances and the
{\it Hipparcos} distances is $1.4 \pm 1.0 \%$, with an rms of 25\%.  Compared
to distances from \citet{santiago15}, our derived distances agree with {\it
Hipparcos} distances slightly better, but are comparable. (For the same data
set, they have distance difference residual and rms of 1.6\% and 26.4\%,
respectively.) \citet{feuillet15} applied a slightly different cut
than ours to get the list of 705 stars. In the list, there are still some stars
whose $\Teff$ are above the cut limits we applied, i.e. $\Teff < 5200
K$. {\bf To} be consistent with other comparison, we excluded these stars from
the list. To exclude stars in possible multiple systems, we also limit our
objects to likely single stars, i.e.  those with ${\tt Soln>10}$ in the
\citet{van07} catalogue.  This process leads to 662 reminding. For these 662
stars, the mean difference between our derived distances and those
from {\it Hipparcos} is -0.6$\pm 1\%$, with an rms of 25\%. This comparison is
shown in Fig.\ref{fig:hip}.

\subsection{Members of Well-Studied Clusters}
\label{sec:cluster}

\begin{figure}
\centering
\includegraphics[width=8.0cm]{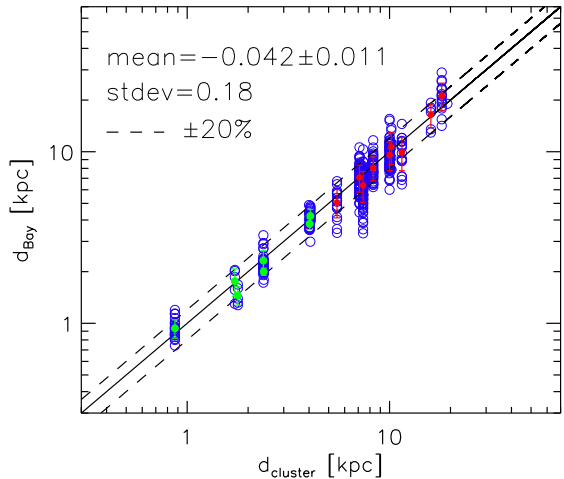}
\caption{ 
Estimated distances to cluster stars plotted against the cluster distances from
the literature, and star members of clusters are identified by
\citet{meszaros13}. The solid line indicates identity, while the dashed-line
shows the distances that differ by 20\%. The mean relative difference and
dispersion are labeled at the top right of the figure. The blue open circles
are for individual stars, while the red and green solid circles indicate the
mean distances derived from stars belonging to global clusters and open
clusters, respectively, and the error bars indicate the dispersion for each of
the clusters.} 
\label{fig:cluster} 
\end{figure}

559 stars in 20 star clusters are used for the APOGEE calibration
\citep{meszaros13}. The identifications of these cluster members are listed in
their Table 4. These members provide a good opportunity to check the
reliability of our derived distances via comparing our distances with
those from a cluster with independent methods.

Among 559 stars used by \citet{meszaros13}, all are members of two open
clusters M35 and Pleiades, but are not included in the new data release SDSS
DR12. For better quality of APOGEE stellar parameters, we also excluded targets
with $\Teff>5200$ K and $\logg > 3.8$ \citep[see][]{holtzman15}.  This leaves
us with 437 stars from 11 global clusters and 7 open clusters, of which 332
stars are in global clusters and 105 are in open clusters. Distances of the 18
clusters are searched from literature, and listed in Table 2 below.  Only
recent measurements, later than 2009, are used. Otherwise, distances in Open
Cluster Catalogues \citep{dias02} and in the Global Cluster Catalogue
\citep{harris96} are used.

Distances of these 437 stars are derived with the Bayesian method as described
in Section 2. As always, identifying cluster members is not trivial work. It is
very possible that a few stars in Table 4 of \citet{meszaros13} are not true
cluster members since their targets include stars with a probability $>50\%$ of
being a cluster member based on their proper motions. Furthermore, it is
possible that cluster members interact each other due to the crowdedness of the
field; such interactions might result in blue stragglers or other outcomes.
Clusters also have a high fraction of binary or multiple systems. The evolution
status of cluster members will change if they experience interactions between
(among) cluster members, and become components of binary or multiple systems.
So, their observational vector, $O=(\Teff, \mh, {\log g}, J, H, Ks)$, will
deviate from their intrinsic ones. As a result, our derived distances of these
members, which are based on their observational vectors, will deviate from the
true cluster distance. This results in outliers in our derived distances to
cluster members. To exclude these outliers and possible ``false'' cluster
members in the target selection, we discarded six stars whose derived distances
are less than one half of the mean distances of other members in the same
cluster, or are greater than 1.8 times the mean distance.

\begin{table*}
\begin{center}
\caption{Cluster Distances from Literature.}
\label{table:cd}
\begin{tabular}{l|l|c|c|c|c|}
\hline
Cluster ID & Name & Distance (kpc) &  Reference  \\
\hline
NGC 4147 &  & $19.3$  & \citet{harris96} \\
NGC 5024 & M53 & $18.1 \pm 0.4$  & \citet{dekany09} \\
NGC 5272 & M3 & $10.0 \pm 0.2$ & \citet{dalessandro13} \\
NGC 5466 &  & $16.0$  & \citet{harris96} \\
NGC 5904 & M5 & $7.4 \pm 0.1$ & \citet{coppola11} \\
NGC 6171 & M107& $5.5$ & \citet{Oconnell11} \\
NGC 6205 & M13  & $7.1 \pm 0.14 $ & \citet{dalessandro13} \\
NGC 6341 & M92 & $8.3 \pm 0.2$ &  \citet{harris96}\\
NGC 6838 &  M71 & $4.0$ & \citet{harris96} \\
NGC 7078  & M15 & $10.2 \pm 0.2$ & \citet{harris96} \\
NGC 7089 & M2 & $11.5$  & \citet{harris96} \\
NGC 188  &          & $1.72 \pm 0.07$ & \citet{wang15}; \citet{jacobson11} \\
NGC 2158 &         & $4.03 \pm 0.13$ & \citet{jacobson11} \\
NGC 2420 &         & $2.39 \pm 0.29$ & \citet{jacobson11} \\
NGC 2682 & M67 & $0.87 \pm 0.15$ & \citet{sarajedini09}; \citet{jacobson11} \\
NGC 6791 &         & $4.06 \pm 0.15$  &  \citet{an15} \\
NGC 6819 &         & $2.38 \pm 0.10$  &    \citet{wu14} \\
NGC 7789 &         & $1.78 \pm  0.06$ & \citet{wu09} \\
\hline
\end{tabular}
\end{center}
\end{table*}

Fig.\ref{fig:cluster} shows a comparison between our derived distances of
individual star versus cluster distances listed in Table 2.  The cluster
distances range from $\sim 0.8$ kpc to $\sim 20$ kpc.  For the whole sample,
the mean difference is $-4.2\pm1.0$ percent, in terms of our derived distances
minus corresponding cluster distances, with a dispersion of 18 percent.
Individually, most clusters have mean differences of $<10\%$ with scatters of
less than 18\%, except for NGC 7789, M5, and M53 which have mean difference
residuals of $-18\%$, $-14\%$, and $+13\%$, respectively.  In groups of open
and global clusters, they have mean differences of $(-0.9\pm1.7)\%$ and
$(-5.2\pm1.9)\%$, and dispersions of 13\% and 19\%.  The group of open clusters
has a better comparison than the global clusters in both the mean difference
and the dispersion. The result is not unexpected for two reasons. First, our
derived distances toward individual cluster members depend on quality of
observational quantities in the observed vector, i.e. $\Teff$, $\mh$, ${\log
g}$, J, H, and Ks (see Section 2). In SDSS DR12, offsets have been applied to
stellar parameters from ASPCAP\citep{garciaperez15} based on calibrations with
clusters, standard stars, and stars in the $Kepler$ field. Among these
observational quantities, ${\log g}$ is the most important one for deriving
distances to individual stars, because surface gravities in SDSS DR 12 are
calibrated via asteroseismic analysis in the $Kepler$ field and stars in the
field have similar metallicities as those of open cluster members. By contrast,
global clusters have much lower $\mh$. Therefore, we expect that stellar
parameters of our open cluster members are better calibrated than those of
global cluster ones. In other words, we expect that our derived distances to
open cluster members are more reliable than those to global cluster stars.
Secondly, globular clusters are much more crowded than open clusters, and they
are usually much farther. So, listed distances of open clusters in Table 2 are
usually more accurate than those of globular clusters.

\citet{santiago15} also derived distances for the same set of cluster members.
Their derived distances are 16.5\% greater than cluster distances with an rms
of 29.9\%. In both the mean difference and dispersion, our distances appear to
agree with cluster distances better than theirs. However, we use cluster
distances from the most recent measurements when they are available whereas
\citet{santiago15} adopted cluster distances from cluster catalogues
\citep{dias02, harris96}. Our adopted new distance measurements may explain
part of our better comparison.

Cluster distances, including those listed in Table 2, are usually derived by
main sequence fitting, white dwarf fitting, period-luminosity relation, proper
motions, etc. Most methods are somewhat model dependent.  As \citet{santiago15}
argued, isochrone-based cluster distances are subject to a zero-point shift.
Each method suffers its own uncertainties. It is not unusual that distances of
the same cluster determined with different methods or by different individuals
differ by 10\%. Therefore, taking into account mutual uncertainties in both
distances, those listed in Table 2 and in our new derived distances, we
conclude that these new derived Bayesian distances agree very well with cluster
distances from different methods.  Based on the fact that our over 100 open
cluster members have a mean difference of $(-0.9\pm1.7)\%$, and a dispersion of
13\%,  we believe that the Bayesian Method can deliver reliable stellar
distances assuming their stellar parameters are well determined.

\subsection{Red clump stars in the APOGEE-RC catalogue}

\label{sec:rc}

\begin{figure*}
\centering
\includegraphics[width=8.1cm]{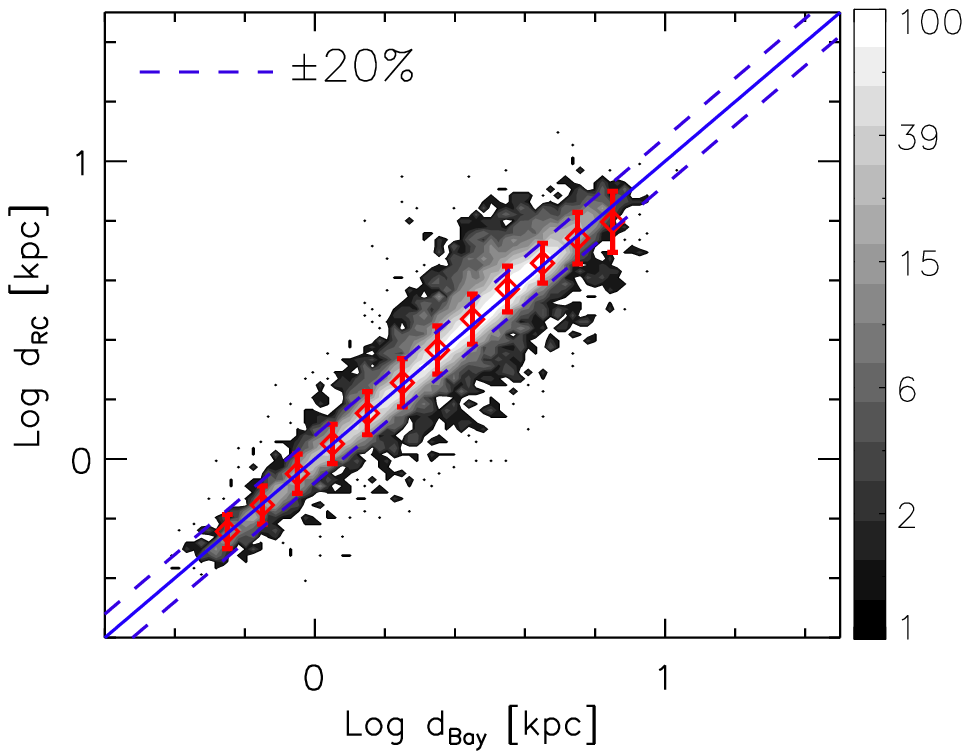}
\includegraphics[width=8.1cm]{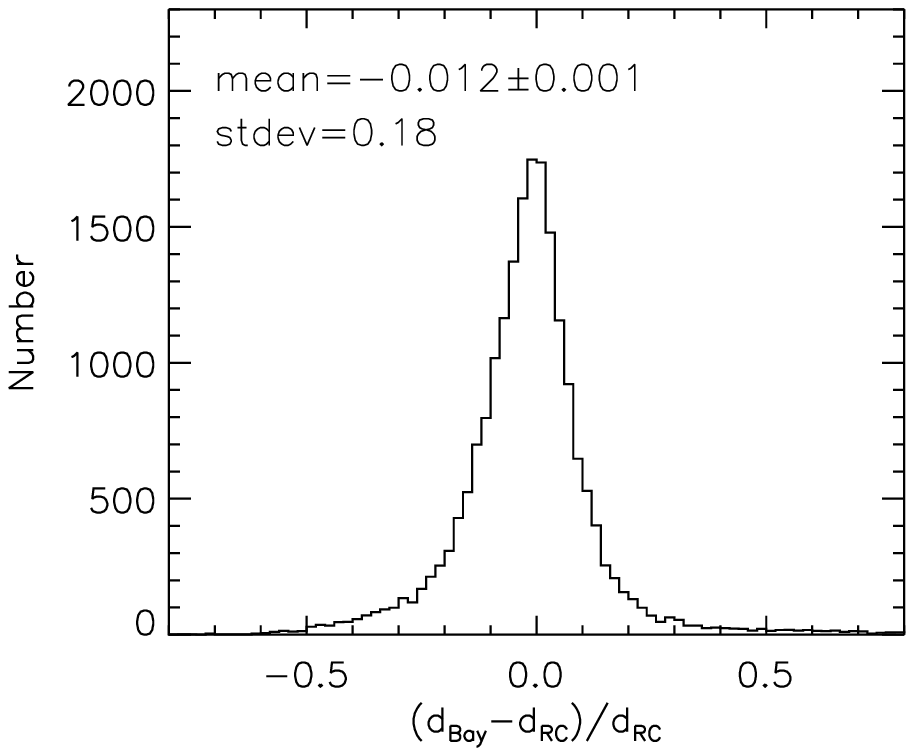}
\caption{ 
Left panel: Comparing distances derived by our Bayesian method (d$_{\rm Bay}$)
to that from \citet{bovy14} for red clump stars (d$_{\rm RC}$). The blue solid
line shows the identity, and two dashed lines indicate distances differ by 20\%
for guidance.  Right panel: the histogram of the distribution of relative
difference with mean value and dispersion being labeled on the top.  The error
of mean value is estimated by the bootstrap method with 1000 samples. }
\label{fig:rc} 
\end{figure*}

The red clump star corresponds to the core-helium-burning stage in the stellar
evolution of low-mass stars, and the luminosity distribution of this type of
star is very narrow, e.g. their absolute magnitude at peak does not depend
strongly on the age and metallicity \citep{groenewegen08}, thus, they have
frequently been used as distance standard candles \citep{williams13}.

Recently, \citet{bovy14} developed a new method for selecting individual likely
red clump stars from spectrophotometric data by combining cut in $(\logg, \Teff, \feh,
[J-K_s]_0)$ and calibrating with high quality seismic $\logg$ data from a
sub-sample of the APOGEE stars with measured oscillation frequencies from the
$Kepler$ \citep{gilliland10} mission. This method results in a sample with high
purity, and it is estimated that the contamination is less than 3.5 percent
\citep{bovy14}. The newly released APOGEE-RC catalogue in SDSS DR12 contains
$19,937$ stars, which is around 20 percent of the whole APOGEE sample. In other
words,  distances towards 80\% of stars in the APOGEE survey cannot be obtained
via the RC distance candle.

\citet{bovy14} derived distances toward their identified RC stars, and
estimated that the distance uncertainty is 5\%\textemdash 10\%.  In
Fig.\ref{fig:rc}, we plot our derived distances versus those from
\citet{bovy14}. The left panel presents the one-to-one correspondence, while
the right panel shows the relative difference distribution. The mean relative
difference in terms of $(d_{\rm ours} - d_{\rm RC})/d_{\rm RC}$ is $-1.2\pm0.1$
percent with an rms of 18\%. It should be noted that this comparison uses all
stars in the APOGEE-RC catalogue which includes some misclassified RGB stars
\citep{bovy14}. RC-distances from \citet{bovy14}  towards these misidentified
RC stars will be off of their true values.

\subsection{Stars with asteroseismic Distance}

\begin{figure*}
\centering
\includegraphics[width=8.1cm]{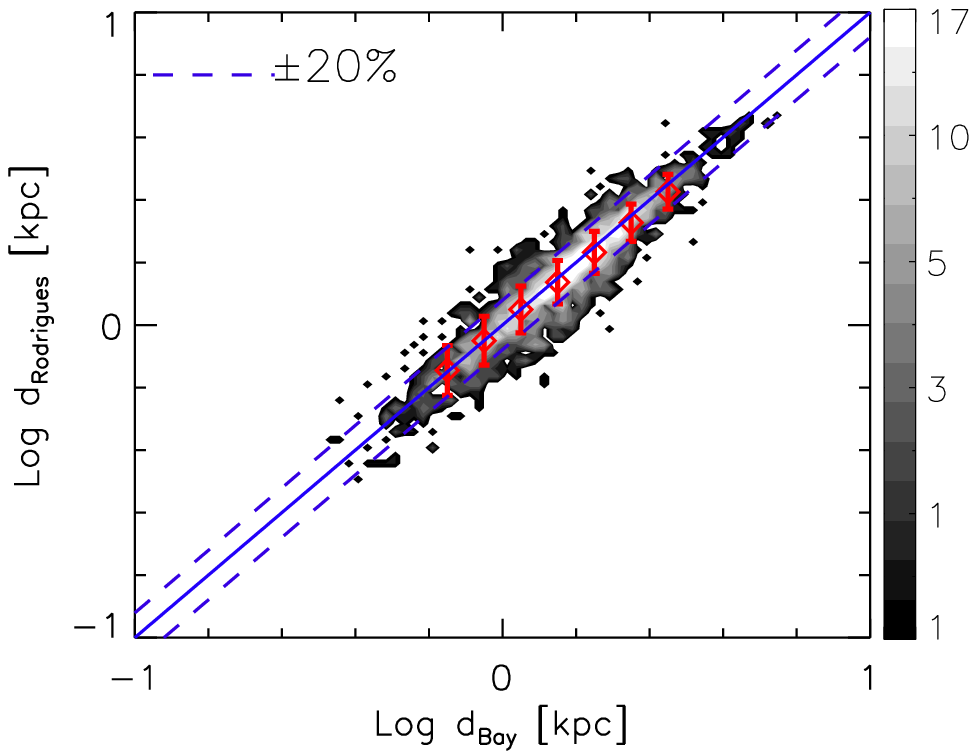}
\includegraphics[width=8.1cm]{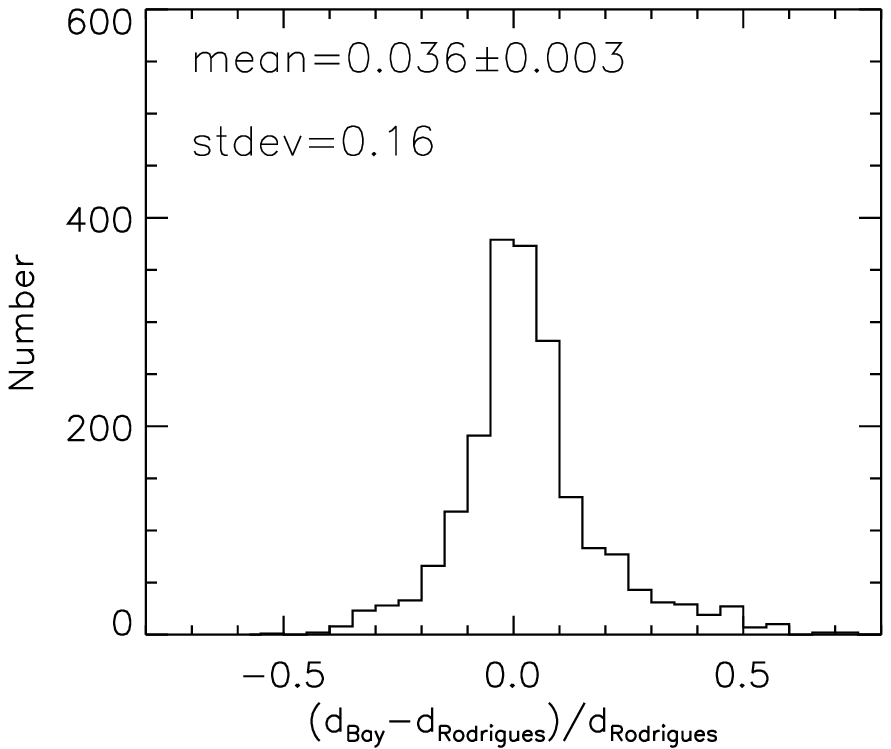}
\caption{ 
Left panel: Comparison of distances derived from our Bayesian method (d$_{\rm
Bay}$) with those from \citet{rodrigues14} (d$_{\rm Rodrigues}$) for stars with
asteroseismologic information available in the $Kepler$ field. The blue solid
line and dashed-lines show the identity and distances that differ by 20\% for
guidance. The red diamonds and error bars are the mean value and dispersion in
each bin respectively. Right panel: the histogram of the relative difference
distribution with mean value and dispersion being labeled on the top. The error
of mean difference is estimated by the bootstrap method with 1000 samples.}
\label{fig:d_rod} 
\end{figure*}

Recently, \citet{rodrigues14} derived the distances and extinctions for 1989
stars in the APOKASC Catalogue \citep{pinsonneault14} with a Bayesian method.
The method took into account the spectroscopic constraints derived from APOKASC
and the asteroseismic parameters from the {\it Kepler} asteroseismic Science
Consortium (KASC). They used the code
PARAM\footnote{http://stev.oapd.inaf.it/param} \citep{desilva06} to estimate
the stellar properties by comparing observation data with those derived from a
stellar model via a Bayesian method. In their work, two steps have been used to
derive the distance and extinction. Firstly, from the observables ($\Teff, \mh,
\deltanu,\numax$), they used PARAM to derive a probability density function
(PDF) for a set of parameters: $\cM, R, \logg$, age($\tau$), and absolute
magnitudes of several passbands $M_{\lambda}$. In the second step, they used
$\Teff$ and $\mh$ derived from spectroscopic observation, as well as the
asteroseismic $\logg$ to derive the PDF of the absolute magnitudes for given
passbands. With PDF for each passband, the joint PDF of the distance modulus
can be obtained after accounting for the extinction.  Distances and extinctions
of 1989 stars were obtained with constraints from asteroseismic $\logg$,
spectroscopic constraints of $\Teff$ and $\mh$ from SDSS DR10 \citep{ahn14},
and photometry constraints from SDSS, 2MASS, and WISE data. Because of the
highly accurate $\logg$ values that are well constrained by their available
asteroseismic information, the distances can be determined with high precision
\citep{rodrigues14}. The internal uncertainties of the derived distance are a
few  percent \citep{rodrigues14}. Such highly accurate distance data are
desirable for the study of Galactic Archaeology. However, the majority of
APOGEE stars  do not have asteroseismic information, so their distances cannot
be derived via this method, i.e. with asteroseismic constraints. Comparing
these accurate distances with those  from our Bayesian method, which does not
require asteroseismic information,  will be useful to quantify the accuracy of
our determination.

The results {\bf are} shown in Fig.\ref{fig:d_rod}. The left panel presents a
one-to-one correspondence, while the right panel indicates the distribution of
the fractional difference. The mean fractional difference is $3.6\pm0.3$ per
cent with a scatter of 16 per cent. Although both \citet{rodrigues14} and we
use Bayesian methods, the two studies use different approaches (see Section 2
and the previous paragraph): different prior functions are adopted,
\citet{rodrigues14} used $\Teff$ and $\mh$ from SDSS DR10 whereas ours are from
SDSS DR12,  \citet{rodrigues14} used photometry information from more
bandpasses than we do.  The difference may also arise from the surface
gravity used in \citet{rodrigues14} and ours, to which the distances are much
sensitive.  Rodrigues et al. (2014) used asteroseismic information directly,
while we adopted the $\logg$ from the calibrated parameters of ASPCAP. The
nominal uncertainty of calibrated $\logg$ from ASPCAP is 0.11 dex, however, the
true error is likely to be dominated by systematics \citep{holtzman15}. It is
possible that the mean difference and the scatter are due to the combination of
the differences in the two methods.  It is worth noting that the median
difference is only $1.7\pm0.3$ per cent in this comparison. Meanwhile, if our
other distance indicator, $1/\ex{\varpi}$, is used, both the mean difference
and the scatter will be smaller, 1.2\% and 15\%.


\begin{figure*}
\centering
\includegraphics[width=8.1cm]{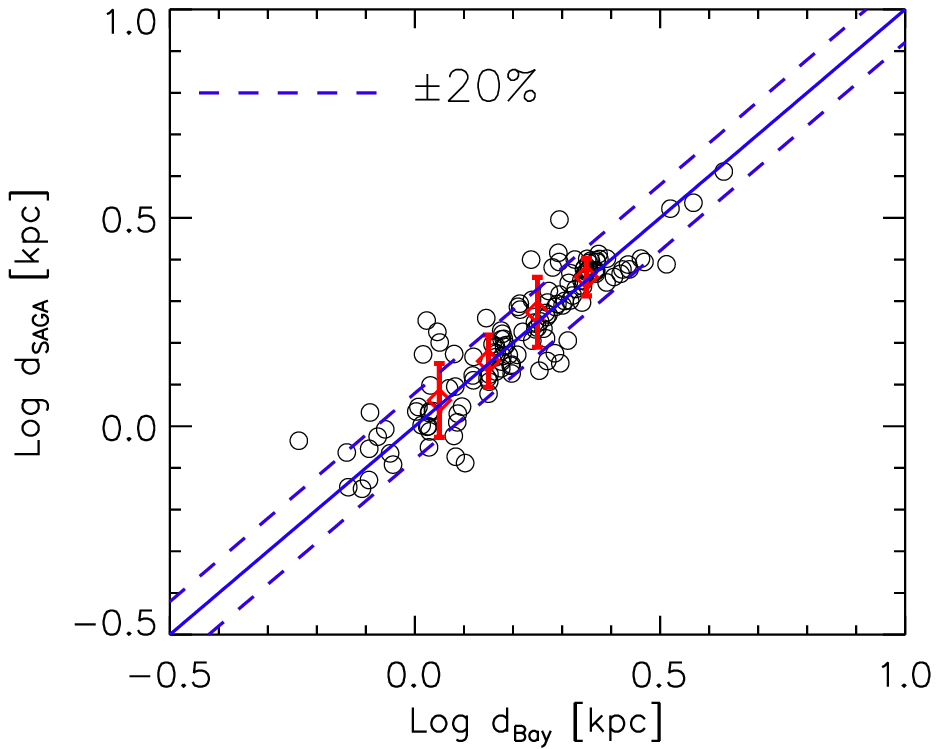}
\includegraphics[width=8.1cm]{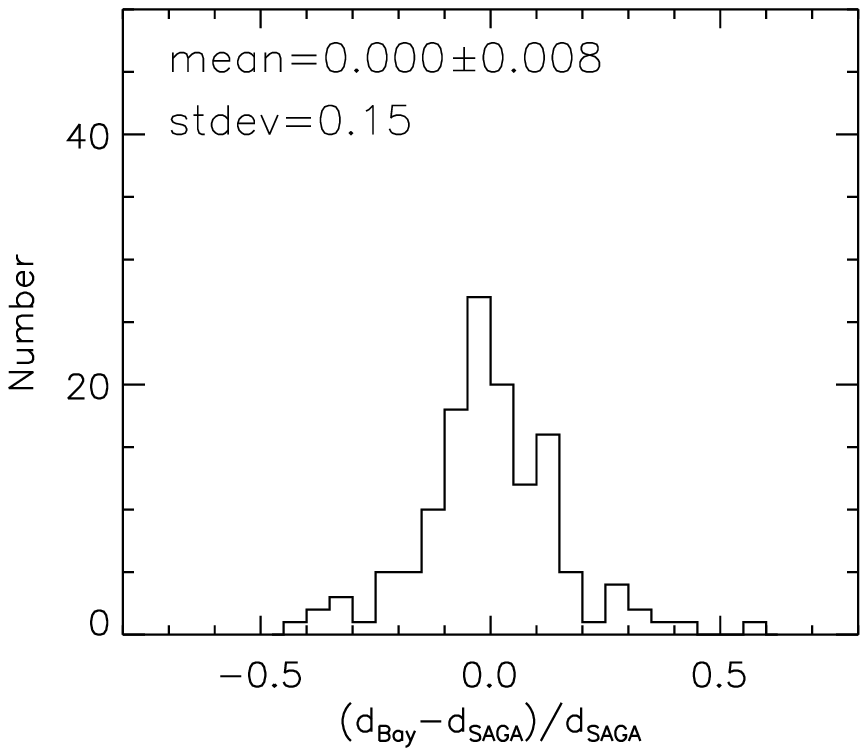}
\caption{ 
Left panel: Comparing distances from this work (d$_{\rm Bay}$) with those from
\citet{casagrande14} (d$_{\rm SAGA}$) for the SAGA catalogue. The black open
circles show individual stars, while the red solid circles and error bars
indicate the mean values and the dispersion. The blue solid line denotes the
identity and dashed-lines show differences at $\pm 20\%$ for guiding eyes.
Right panel: the histogram of the relative difference with mean value and
dispersion being labeled on the top.  The error of the mean value is estimated
by the bootstrap method with 1000 samples. } 
\label{fig:d_saga} 
\end{figure*}

\citet{casagrande14} released the first results from the ongoing Str\"{o}mgren
survey for Asteroseismology and Galactic Archaeology (SAGA) in the $Kepler$
field. They derived the stellar parameters by coupling of classic and
asteroseismic parameters iteratively and self-consistently: e.g. effective
temperatures from IRFM, metallicities from Str\"{o}mgren indices, and masses
and radii from seismology. This data set is limited by the {\it Kepler} and
Str\"{o}mgren surveys,  so only a small strip of sky coverage centered at a
Galactic longitude of 74$\degr$ and covering latitude from about $8\degr$ to
$20\degr$ is available.  Their first release only includes $\sim 1000$ stars,
but the typical precision of their derived distances is high, at a few percent
\citep{casagrande14}. We cross-matched \citet{casagrande14}'s catalogue with
the APOGEE stars, and found 135 matches. We compare our derived distances with
those from the catalogue. The results are presented in Fig.\ref{fig:d_saga}.
Surprisingly, There is nearly no systematical offset, and the dispersion is 15
percent.

No systematic offset between the two sets of derived distances is a somewhat
surprising result because there are differences in the underlying methods, in
isochrones used, and spectroscopic stellar parameters used.
\citet{casagrande14} derived the stellar parameters in a different way than the
APOGEE pipeline, and there are systematic differences in the derived parameters
from these two different methods, i.e. their $\Teff$s are about 90 K hotter
than those from ASPCAP,  and [M/H]s are  $\sim 0.14$ dex lower. The differences
in $\Teff$ and $\mh$ could introduce a $\sim 2\%$ systematic difference in
derived distances \citep{rodrigues14}.  It is possible that the difference
introduced by $\Teff$ and $\mh$ is offset by ones due to other differences,
such as methods and isochrones.

\section{Extinction}
\label{sec:extinction}

\subsection{Comparison with extinction from \citet{rodrigues14}}

\begin{figure*}
\centering
\includegraphics[width=8.1cm]{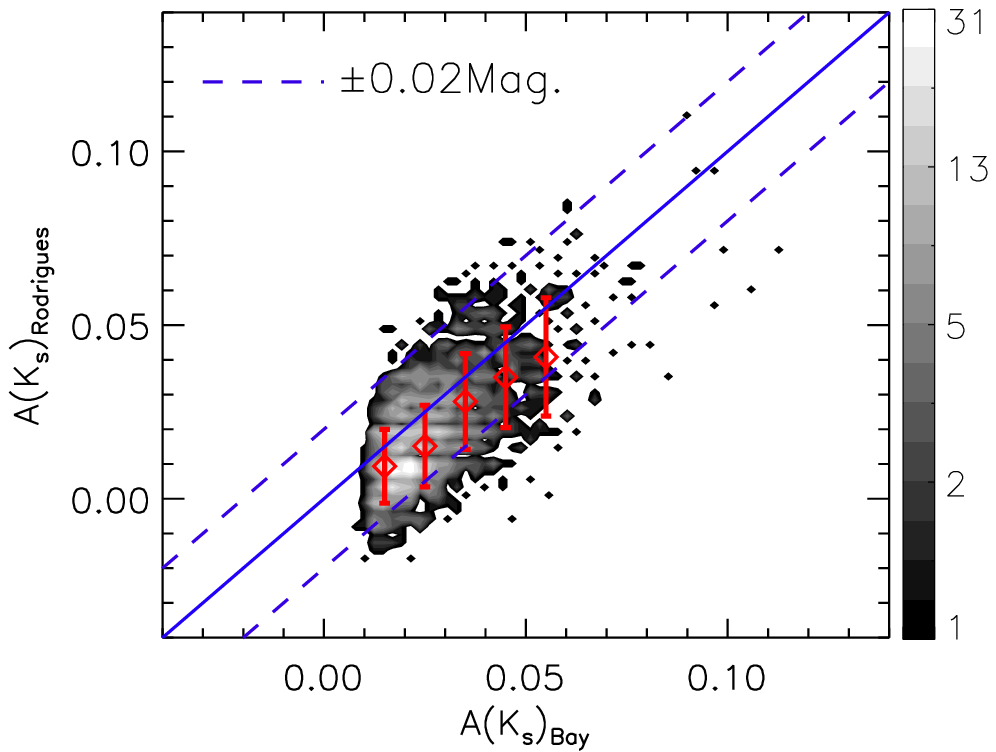}
\includegraphics[width=8.1cm]{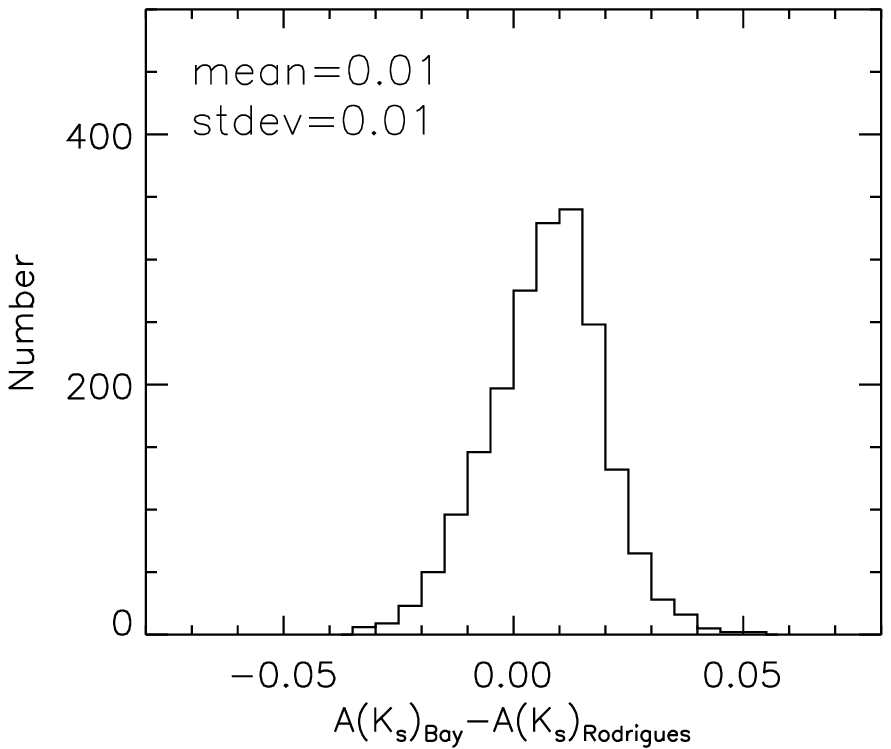}
\caption{ 
Left panel: compares extinction in $K_s$ band derived from our Bayesian method,
A(K$_s$)$_{\rm Bay}$, with that derived by \citet{rodrigues14}, A$(K_s)_{\rm
Rodrigues}$, for stars in the $Kepler$ fields. The solid blue line shows the
identity, while the dashed blue lines indicate the extinction differs by 0.02
Mag. The red symbols show the mean value, while the error bars indicate the
dispersion. Right panel: the histogram of extinction difference with mean and
dispersion being labeled on the top.  } 
\label{fig:Ak_rod} 
\end{figure*}

\citet{rodrigues14} obtained distances and extinctions of stars with the
asteroseismic constraints. It is believed that these extinctions are very
accurate, thus, it is instructive to compare our extinctions with theirs.
Fig.\ref{fig:Ak_rod} compares our results with those of \citet{rodrigues14}.
The left panel presents a one-to-one correspondence for each star, while the
histogram of the extinction difference in the K$_s$-band is shown in the right
panel.  The plots show that the two sets of extinctions agree very well. The
mean difference is only 0.01 Mag.. It is worth noting that, unlike
\citet{rodrigues14}, our Bayesian method does not return any (nonphysical)
negative $A(K_s)$.

The {\it Kepler} field is a low extinction region, so the extinctions towards
most stars are less than 0.07 Mag. in $K_s-$band, which corresponds to 0.6 Mag.
in $A_V$.  This value is much lower than that in the disk region, where most
APOGEE targets are located. Thus, in the following section we compare our
extinction with those from other methods.

\subsection{Comparison with Extinctions from RJCE and from the Two-dimensional Extinction Map}

\begin{figure*}
\centering
\includegraphics[width=8.1cm]{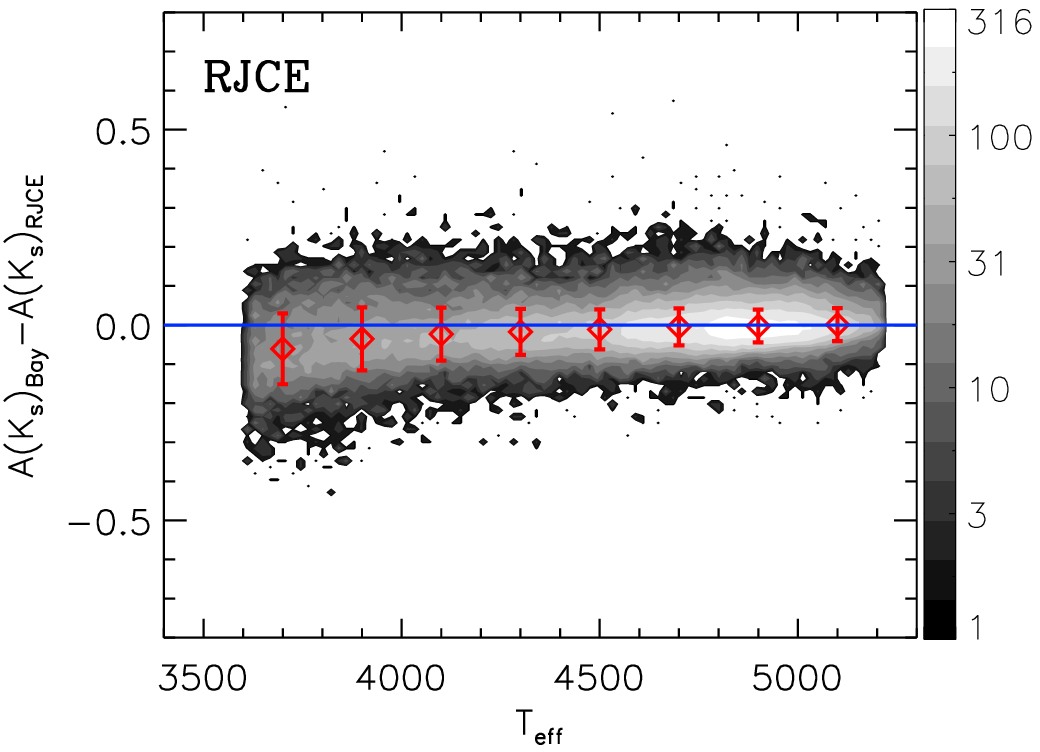}
\includegraphics[width=8.1cm]{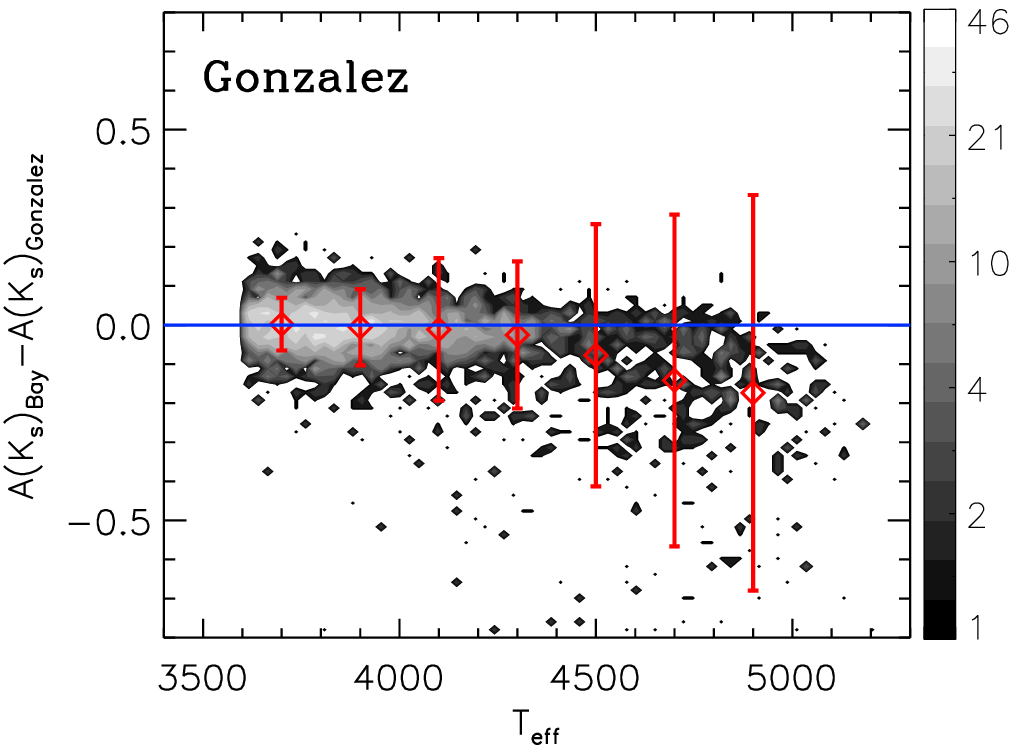}
\includegraphics[width=8.1cm]{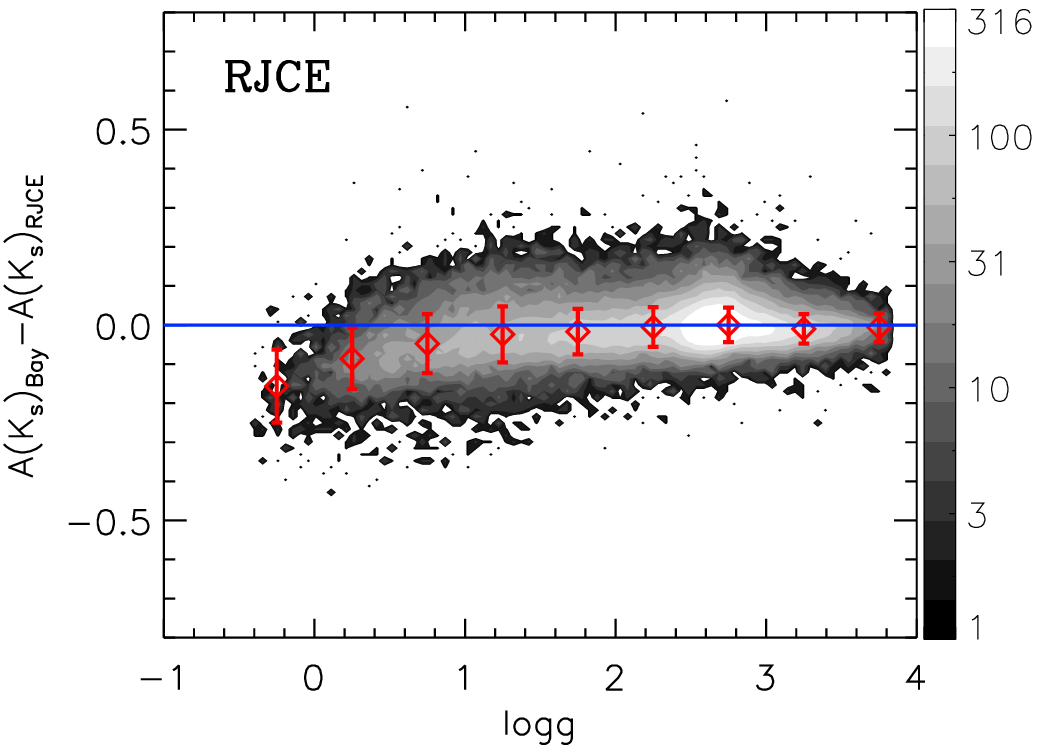}
\includegraphics[width=8.1cm]{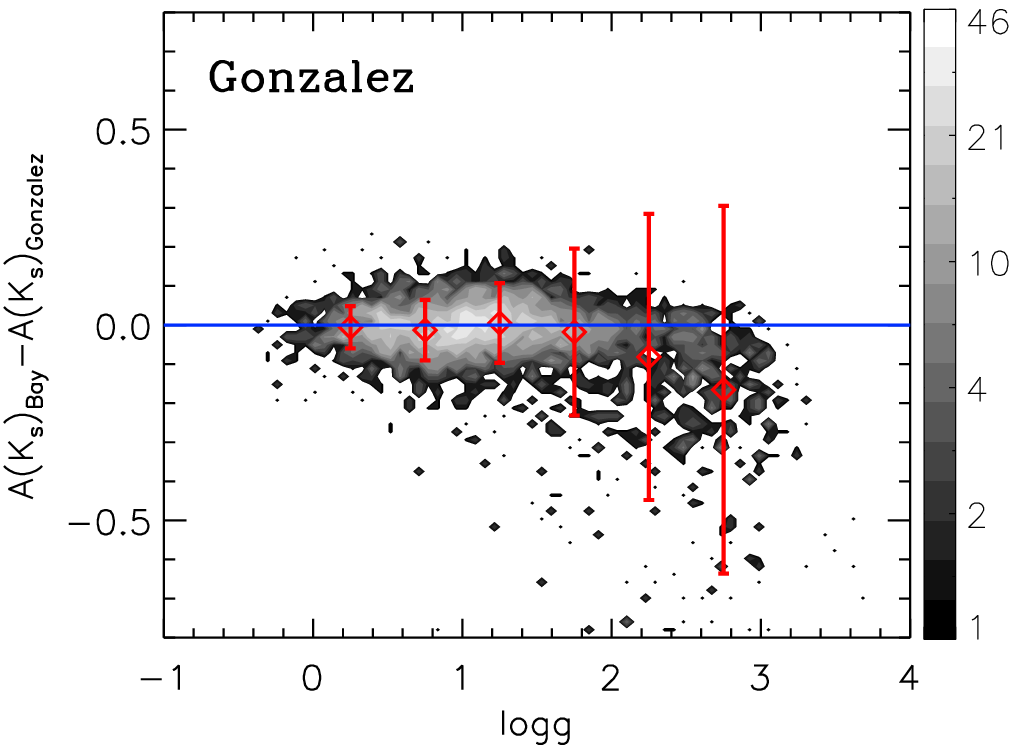}
\includegraphics[width=8.1cm]{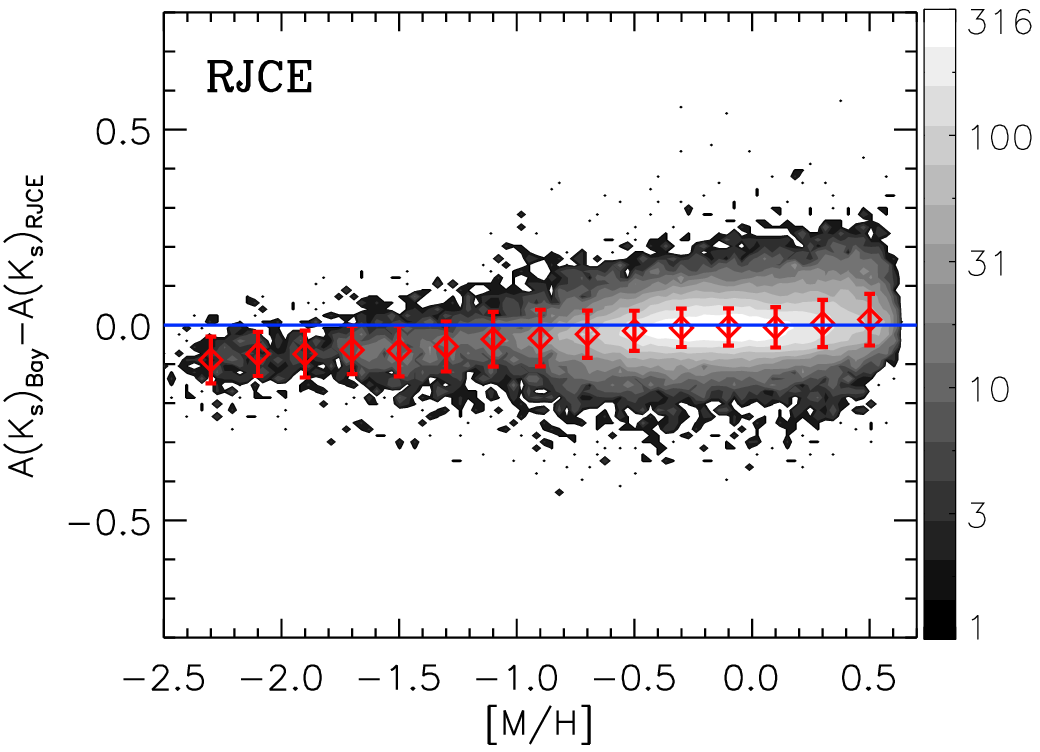}
\includegraphics[width=8.1cm]{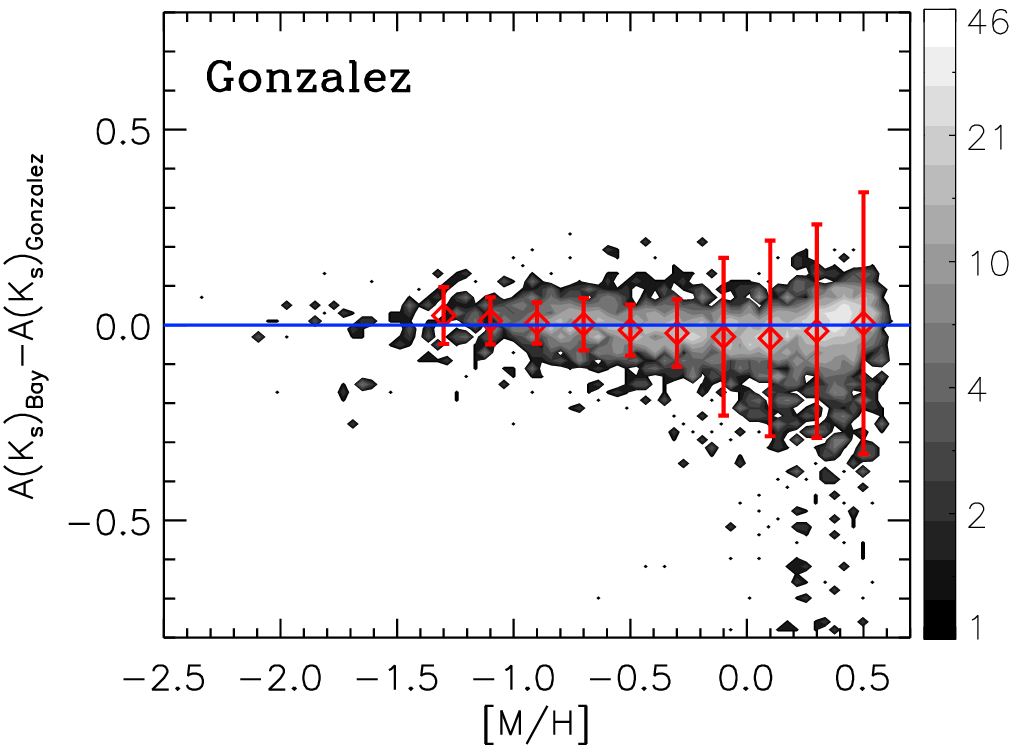}
\caption{ 
Comparisons of our derived extinction, $A(K_s)_{\rm Bay}$, to those from RJCE,
$A(K_s)_{\rm RJCE}$ (left column), and to G12 $A(K_s)_{\rm Gonzalez}$ (right
column) as a function of $\Teff, \logg$, and $\mh$. The blue solid lines show
the zero. The red symbols show the median values, and the error bar indicates
the standard deviation for every bin. The extinction difference shows clear
trends with $\Teff, \logg$, and $\mh$ in the comparison with RJCE, while there
is no systematic offset in the comparison with G12 $A(K_s)_{\rm Gonzalez}$, see
text for details} 
\label{fig:Ak_param} 
\end{figure*}

To derive the extinction corrections, the Rayleigh Jeans Color Excess (RJCE)
method has been adopted in the APOGEE targeting \citep{majewski11}. RJCE
calculates reddening values on a star-by-star basis using a combination of
near- and mid-IR photometry. This method assumes that all stars have a specific
color in $(H-[4.5{\mu}{\rm m}])_0$.

\begin{equation} \label{equ:rjce}
\displaystyle A(K_s) = 0.918 (H-[4.5{\mu}{\rm m}]-(H-[4.5{\mu}{\rm m}])_0),
\end{equation}

The 4.5$\mu$m photometric data are either from {\it Spitzer}-IRAC data
\citep{werner04,fazio04} or {\it WISE} surveys \citep{cutri13}.
\citet{majewski11} estimated an approximate RJCE extinction uncertainty of $\la
0.11$ Mag. for a typical star. The RJCE method has been widely used for
individual stellar extinction corrections in APOGEE target selection. We
extracted extinction values for stars with  the $ak\_targ\_method$ flag set to
``RJCE" in SDSS DR 12, and compare with our derived extinctions for the same
stars.

We also compared our extinctions with those from the two-dimensional extinction
map of \citet[][G12]{gonzalez12}, which was constructed based on measuring the
mean $J-K_s$ color of the red clump stars in the bulge area. This extinction
map has a resolution limit of 2$\arcmin$. Using the BEAM calculator Web
page\footnote{http://mill.astro.puc.cl/BEAM/calculator.php}, we retrieved the
extinction for each star in the map with the criterion of being the closest
with 2$\arcmin$ of its position.

Fig.\ref{fig:Ak_param} shows the extinction difference between ours and those
from RJCE (left column), and those from \citet{gonzalez12} as functions of the
stellar atmospheric parameters, $\Teff, \logg, \mh$. In general, the extinction
derived from RJCE and \citet{gonzalez12} agree well with ours, with a
discrepancy of $A(K_s) \la 0.2$ Mag. However, we note that, for cool giants,
the extinction from RJCE is systematically higher than ours; this trend, as
also seen in \citet{wang2015}, can be explained by the dependence of the RJCE
method on the spectral type, which is clearly shown in Fig.1 of
\citet{majewski11}.  At low temperature the intrinsic color of
$(H-[4.5{\mu}m])_0$ increases with decreasing temperature, which results in
overestimating the extinction, $A(K_s)_{\rm RJCE}$. In the metallicity panel,
$A(K_s)_{\rm RJCE}$ seems overestimated for stars with $\mh<-1$. These trends
are not manifested in the comparison between our extinctions and those from
\citet{gonzalez12}. Actually,  the $A(K_s)_{\rm RJCE}$ overestimation has been
previously noticed by \citet{zasowski13} and \citet{schultheis14a}.

The extinction from \citet{gonzalez12} does not systematically deviate from our
results. The comparison does indicate some outliers, which can be explained by
the low resolution map of \citet{gonzalez12}. There are some stars with large
$\logg$ and high $\Teff$, whose extinctions are overestimated by
\citet{gonzalez12}.  This confirms  \citet{schultheis14a}'s finding that the
extinction to some of nearby stars have been systematically overestimated in
the two-dimensional extinction map.

\begin{figure*}
\centering
\includegraphics[width=8.1cm]{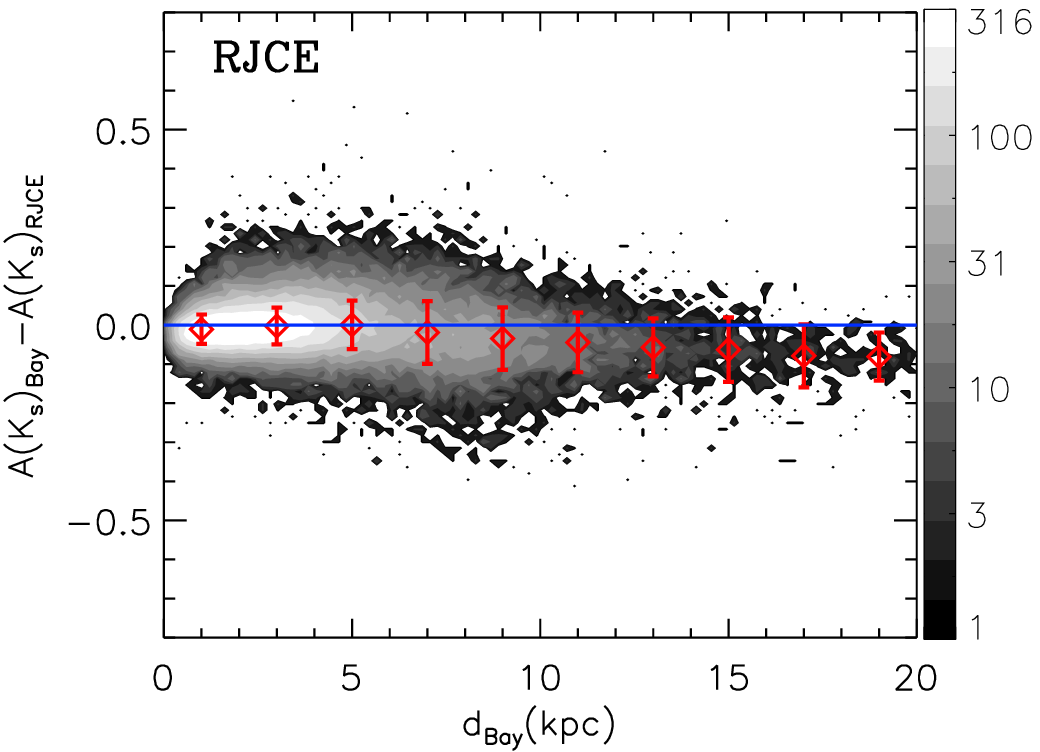}
\includegraphics[width=8.1cm]{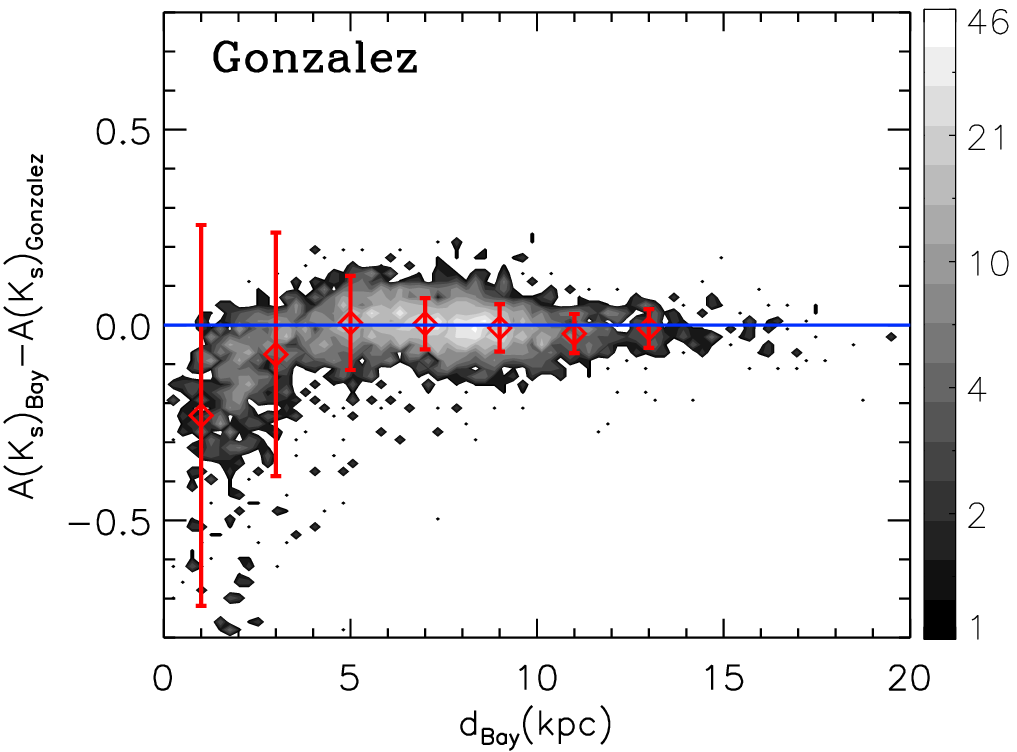}
\includegraphics[width=8.1cm]{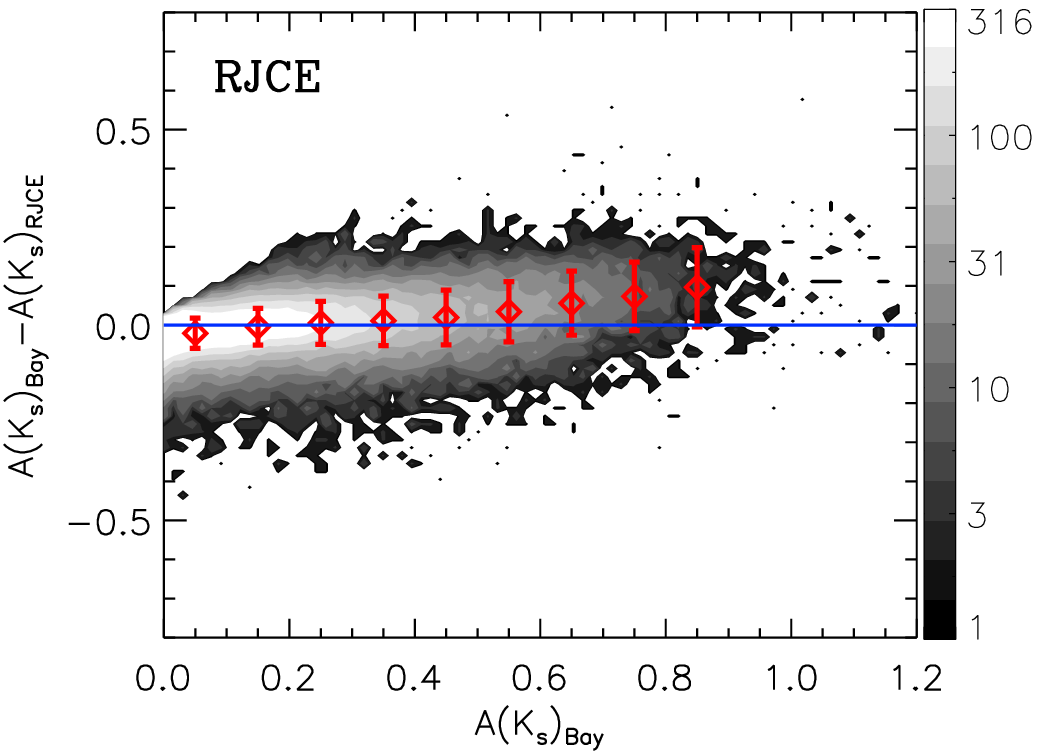}
\includegraphics[width=8.1cm]{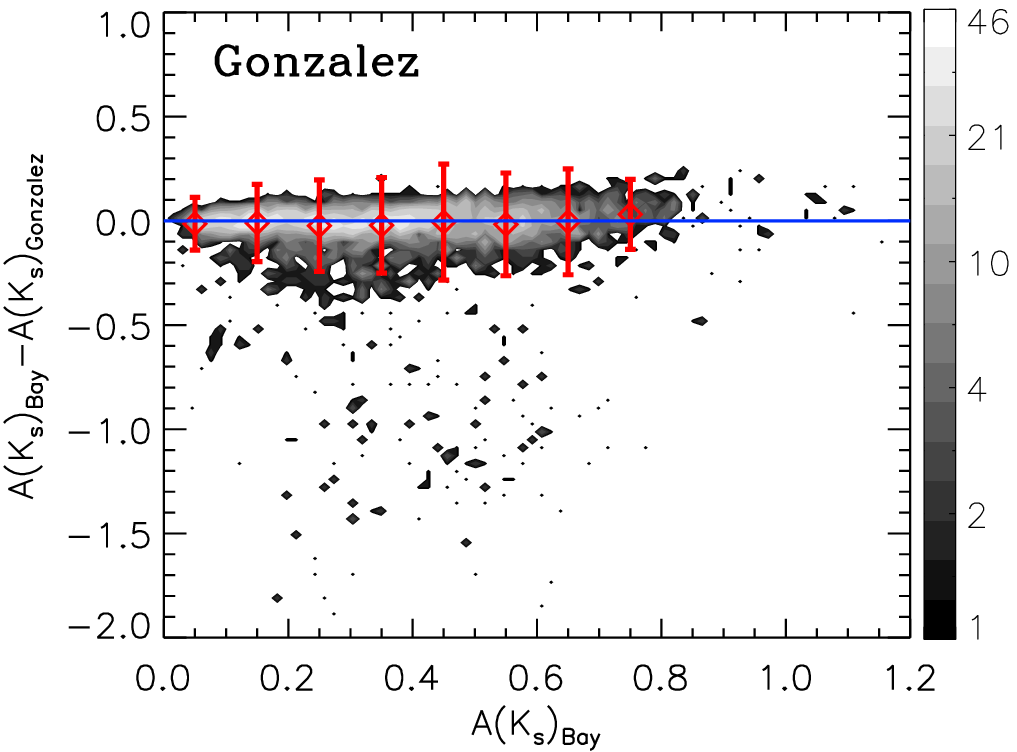}
\caption{
Comparisons of our derived extinction, $A(K_s)_{\rm Bay}$, with those of RJCE
$A(K_s)_{\rm RJCE}$ (left column), and G12, $A(K_s)_{\rm Gonzalez}$ (right
column) as functions of $A(K_s)_{\rm Bay}$ and our derived distance d$_{\rm
Bay}$. The red symbols indicate the median values, while error bars show the
dispersion in bins.}
\label{fig:Ak_dAk}
\end{figure*}

The extinction difference as functions of distance and extinction itself is
presented in Fig.\ref{fig:Ak_dAk}. The top left panel shows that there is a
systematical offset when $d > 9$ kpc.  As shown in the left panel of the second
row in Fig.\ref{fig:Ak_param} of \citet{majewski11}, the RJCE method tends to
overestimate extinctions of giants with low $\logg$.  Because the fraction of
low $\logg$ giants increases with increasing distance in the comparison
samples, a systematic offset emerges at greater distance due to RJCE tending to
overestimate extinctions for objects with low $\log g$. When comparing our
derived extinctions with those from \citet{gonzalez12}, two sets of data agree
very well, with no any noticeable systematic offset for stars with $d > 4$ kpc.
However, when $d<4$ kpc, extinctions from \citet{gonzalez12} are consistently
lower than ours.  Figure 9 of \citet{schultheis14a} shows the same trend.  The
systematic offset may be attributed to the average distance of the tracers that
were used to create the extinction map in \citet{gonzalez12} being farther away
than these relatively close objects. The map overestimated extinctions of these
foreground stars.

In the bottom row of Fig.\ref{fig:Ak_dAk}, we plot the extinction difference
versus extinction itself.  A systematical offset can be seen in the heavy
extinction region, i.e. $A(K_s)>0.5$, where extinction of RJCE  is about $\sim
0.1$ Mag. lower than ours.  A $\sim 0.1$ difference in the extinction
corresponds to about a $5\%$ difference in distance.  On the other hand, the
extinction from \citet{gonzalez12} agrees very well with ours till high
extinction, $A(K_s) \sim 0.8$.  No systematical offset is seen in the
comparison. Above extinction of $A(K_s) \sim 0.8$, we cannot make any
meaningful comparison due to a small sample size.

\subsection{Comparison with Three-Dimensional Maps and Models}

\begin{figure*}
\centering
\includegraphics[width=8.1cm]{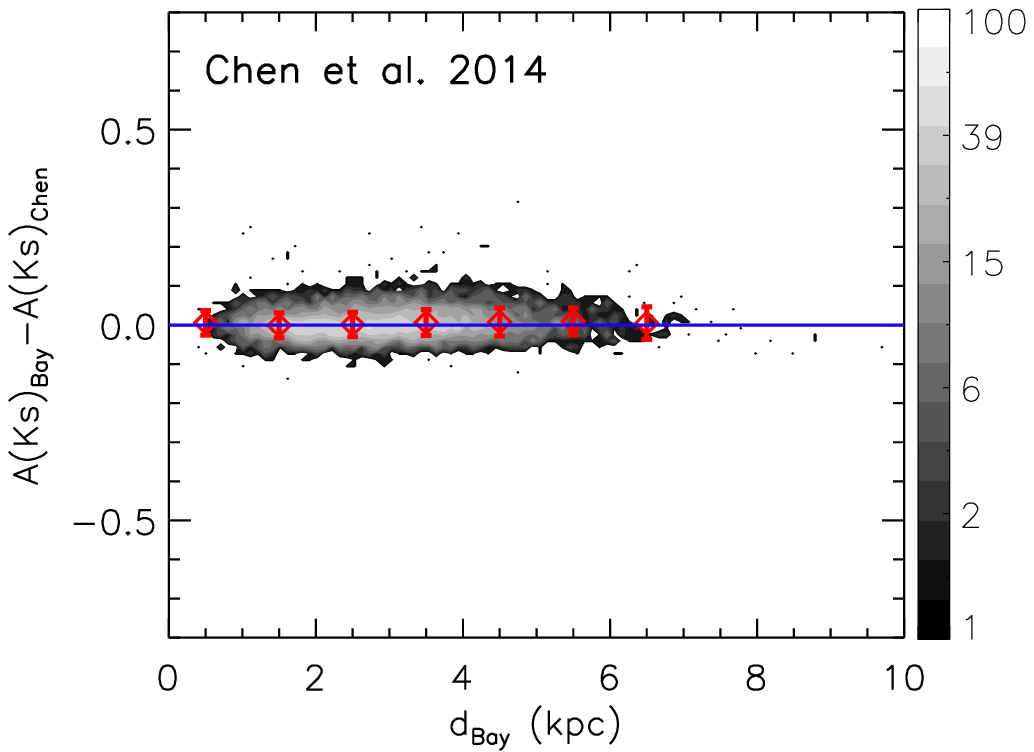}
\includegraphics[width=8.1cm]{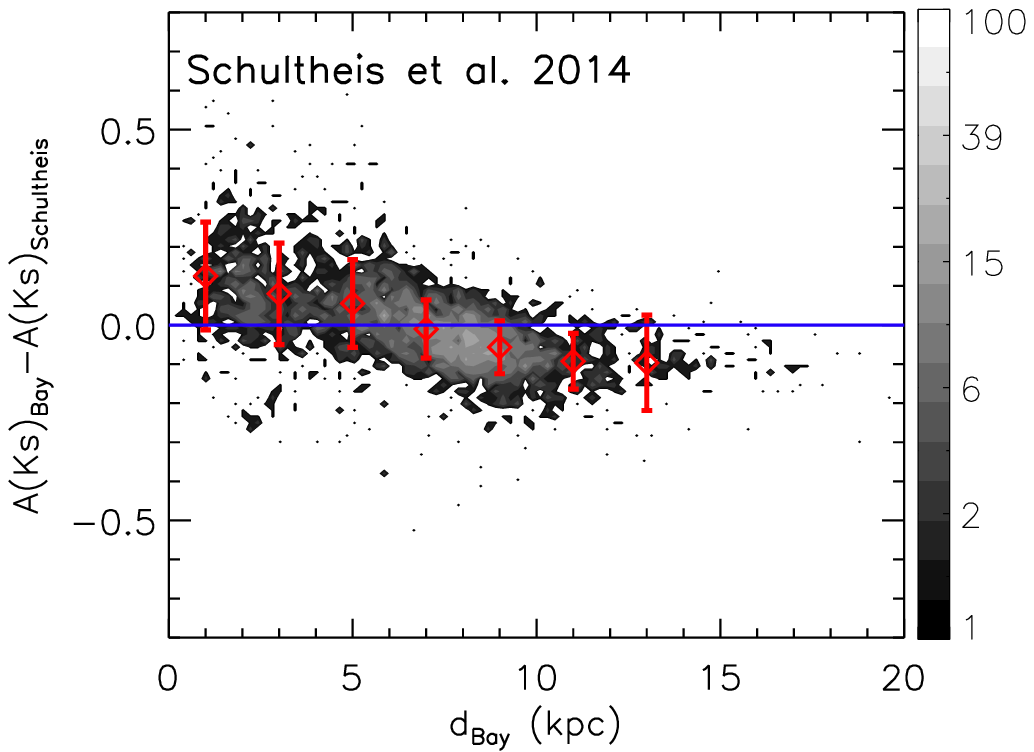}
\includegraphics[width=8.1cm]{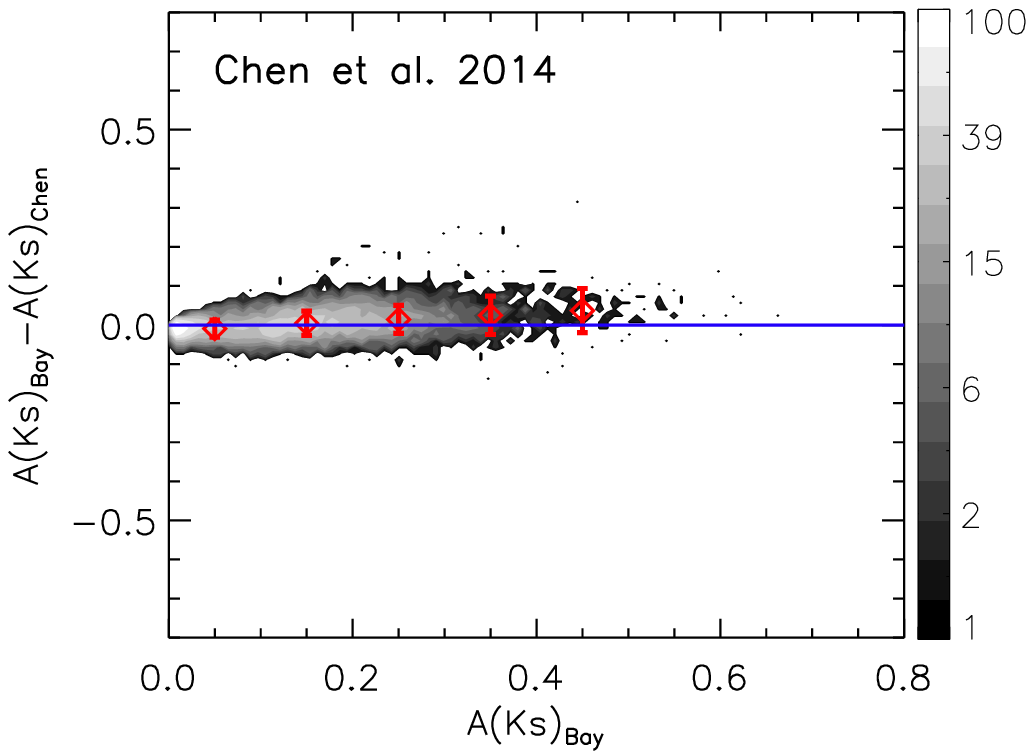}
\includegraphics[width=8.1cm]{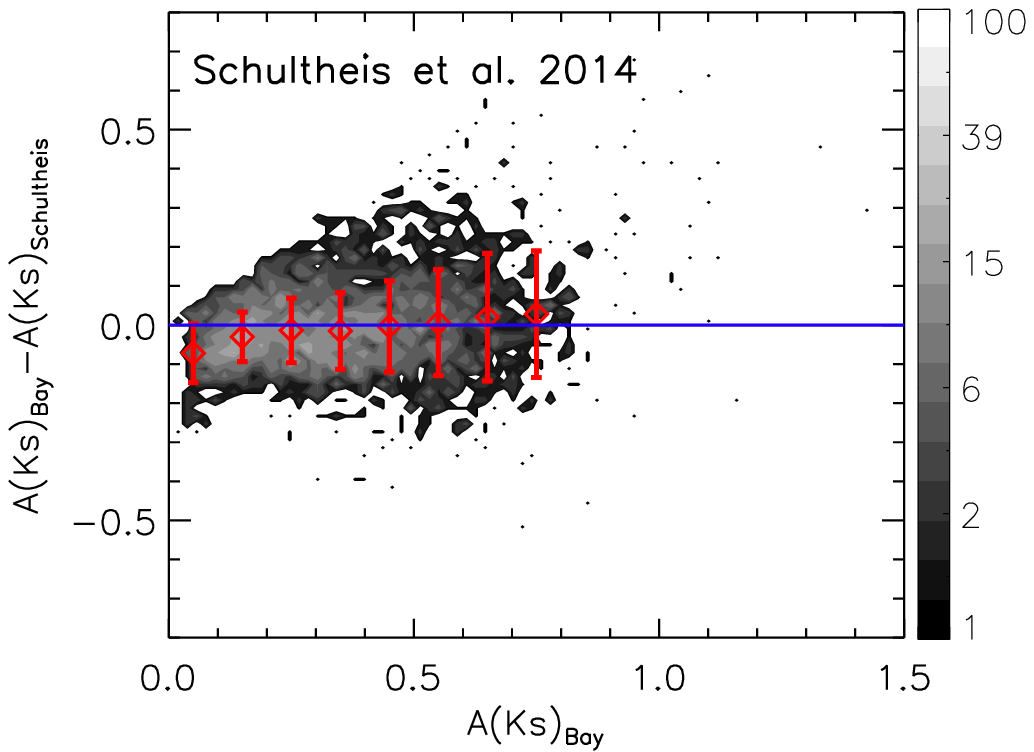}
\caption{
Extinction difference between A(Ks) derived from our Bayesian method,
A(Ks)$_{\rm Bay}$ and that from the 3D extinction map as function of distance and
extinction.  The 3D extinction maps are from \citet{chen14}, A(Ks)$_{\rm Chen}$
and the map of \citet{schultheis14b}, A(Ks)$_{\rm Schultheis}$. }
\label{fig:Ak_3dmap}
\end{figure*}

\begin{figure*}
\centering
\includegraphics[width=17.0cm]{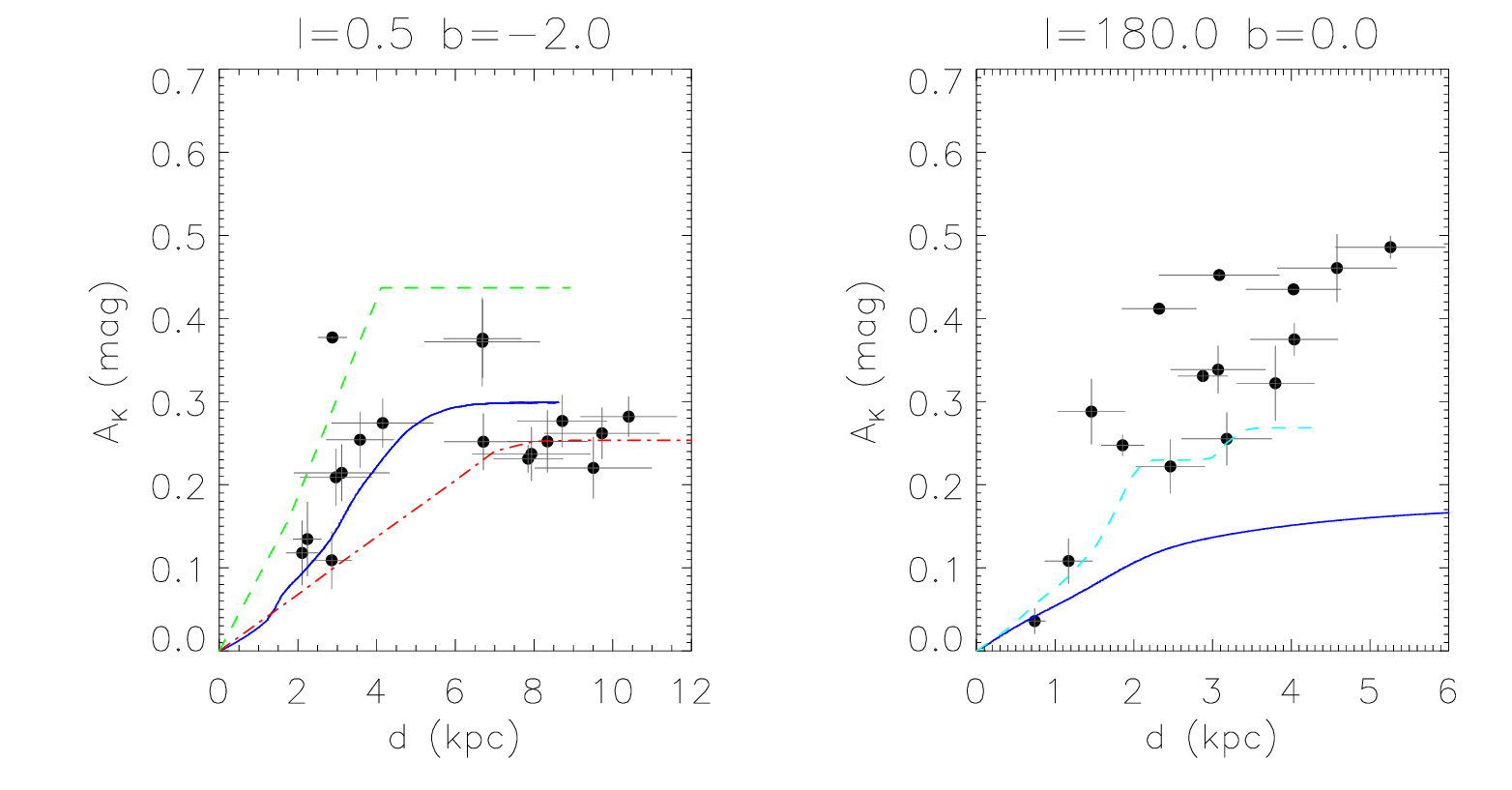}
\caption{
Comparison of 3D extinction map with models in the literature for two fields, one is
toward the Galactic bulge region with $l=0.5$ and $b=-2.0$, and the other one toward the
anti-galactic center region with $ l=180.0$ and $b=0.0$. Dots with error bars
indicate result of this work. Left panel: blue line is from \citet{drimmel03},
red dot-dashed line: \citet{schultheis14b}, green dashed line: \citet{marshall06}  Right panel:
blue line is from \citet{drimmel03}, cyan dashed line: \citet{chen14}.  }
\label{fig:Ak_3dmodel}
\end{figure*}

In this section we compare our derived extinction with the three-dimensional extinction
distributions. Several data sets have been used in this comparison.
\begin{enumerate}

\item The 3D extinction map of \citet{chen14}, which is a
three-dimensional extinction map toward Galactic Anticenter area covering
Galactic longitude $140{\degr}<l<240{\degr}$ deg and latitude
$-60{\degr}<b<60{\degr}$ deg. By combining photometric data from optical to the
near-infrared, they have built a multiband photometric stellar sample of about
30 million stars, and applied spectral energy distribution (SED) fitting to
derive this three-dimensional extinction map with spatial angular resolution
between 3$\arcmin$ and 9$\arcmin$.

\item The map of \citet{schultheis14b}. By using data from the VISTA
Variables in the Via Lactea survey together with the Besançon stellar
population synthesis model of the Galaxy\citep{robin03,robin12,chen13},
\citet{schultheis14b} built a 3D extinction map that covers the Galactic bulge
region with $-10{\degr}<l<10{\degr}$ and $-10{\degr}<b<5{\degr}$, which has a
resolution of $6\arcmin\times6\arcmin$.

\item The \citet{drimmel03} model. Based on the dust distribution model
of \citet{drimmel01}, \citet{drimmel03} built a 3D galactic extinction model.
This model is constructed by fitting the far- and near-IR data from the
$COBE/DIRBE$ instrument, with the pixel size of $COBE$ being about
$21\arcmin\times21\arcmin$.

\item The \citet{marshall06} model, which is built by combining 2MASS
near infrared data with the stellar population synthesis model of Besan\c{c}on
\citep{robin03}, covering the region of $|l| \le 90^\circ$ and $|b| \le
10^\circ$. This map has a spatial resolution of $15\arcmin$. 

\end{enumerate}

Fig.\ref{fig:Ak_3dmap} shows the results comparing our Bayesian
extinction, A(Ks)$_{\rm Bay}$ with those derived from two 3D extinction maps,
i.e.  \citet{chen14} and \citet{schultheis14b}, in which the differences of
extinction are plotted against with distance and extinction values,
respectively.  We cross-match APOGEE stars with the \citet{chen14} data set by
searching the radius within $3\arcsec$. To match with \citet{schultheis14b}
map, we determine the extinction of stars by binning the APOGEE stars to a 6'
resolution, and determine their extinction with the linear interpolation of
(distance, A(Ks)) from the 3D extinction map. For consistency, all the maps
have been re-scaled to A(Ks) using the \citet{ccm89} extinction law.  The map
of \citet{chen14} is built toward the anti-galactic central region, while the
\citet{schultheis14b} map is toward the bulge region.  Our Bayesian extinction
is well correlated with that of \citet{chen14}, such that the mean difference
and dispersion are the smallest among these comparisons. There is nearly no
systematic offset in the plot of the difference versus distance (the top left
panel of Fig.\ref{fig:Ak_3dmap}), but in the plot of difference against
extinction (the bottom left panel of Fig.\ref{fig:Ak_3dmap}) there is a
systematic trend that our Bayesian extinction tends to be larger than that
in map of \citet{chen14} at large extinction.

In the comparison with the 3D extinction map of \citet{schultheis14b},
there is a systematic trend that our derived extinction is larger than that
from \citet{schultheis14b} in a nearby region $(d<7$ kpc$)$, and smaller for
$d>7$ kpc.  
The reason of the systematic difference in the nearby region is not clear,
but it could not come from the saturated magnitude by nearby stars in the data set 
of VVV photometry, because the VVV data set is combined with 2MASS counterparts 
for brighter source.
The difference in the distant region may
come from the sample selection effect of the APOGEE survey, in which stars with
heavy extinction are not observed.  However, the samples used in
\citet{schultheis14b} are complete samples, which have no selection effect.
The scatter of difference in extinction is larger than that compared with
\citet{chen14}. This may be partly caused by the high extinction errors of the
high extinction in the Galactic Bulge, the errors of the distance and the large
variations of extinction in the bins of the 3D extinction map.  However, in the
comparison with data set of \citet{chen14}, it is a star to star comparison.

Having calculated the distance and extinction to each of the APOGEE stars, we can
easily build our 3D extinction map and compare it with the 3D extinction model.
Fig.\ref{fig:Ak_3dmodel} compares our results with the three 3D extinction map
toward two different fields. The left panel of Fig.\ref{fig:Ak_3dmodel} shows
results toward bulge region $(l=0.5,b=-2.0)$, while the right panel of
Fig.\ref{fig:Ak_3dmodel} presents results to the anti-galactic central region
$(l=180.0,b=0.0)$.  We queried those 3D extinction maps for the extinction at
the position (l, b) of each of our APOGEE bulge stars, within a FOV of 0.25
deg$^2$. This FOV was chosen to contain a sufficient number of APOGEE stars
with accurate stellar parameters at each spatial position. Because the
extinction maps have different spatial resolutions, we took the median value
around the center position of each 0.25 deg$^2$ field.  For the field of
$(l=0.5,b=-2.0)$, we see that the \citet{schultheis14b} model is best matched
with our result for region $d>6$ kpc, while for nearby region $d<6$ kpc the map
of \citet{drimmel03} better represents the result.  The model of
\citet{marshall06} predicts a rather steep slope for $d<4$ kpc.  In the comparison
with anti-galactic central field $(l=180.0,d=0.0)$, the maps of \citet{chen14}
and the model of \citet{drimmel03} can give consistent results with ours for
$d<1$ kpc, while for the distant field at $d>1$ kpc, both maps give lower extinction
values comparing to our results.

\section{Summary}
\label{sec:summary}

\begin{table*}
\begin{center}
\caption{The derived spectrophotometric distances and extinctions with our Bayesian method for APOGEE stars.
The full catalog contains 101726 stars and is only available in electronic form.}
\label{table:catalogue}
\begin{tabular}{c|c|c|c|c|c|c|c|c|}
\hline
apogee$\_$id &  R.A. & Decl. & $\ex{d}$ & ${\sigma}_{\ex{d}}$ & $\ex{\varpi}$ & ${\sigma}_{\ex{\varpi}}$& $A_{\rm Ks}$ & ${\sigma}_{\rm Aks}$ \\
             &  (deg)& (deg) &  (kpc)   &  (kpc)              &  (mas)       &   (mas)                 & (Mag.)       &(Mag.)  \\
\hline
2M12111619+8655554 & 182.8174591 & 86.9320755 & 0.2883 & 0.0486 & 3.5608 & 0.5555 & 0.0587 & 0.0201 \\
2M12055981-0307535 & 181.4992218 & -3.1315529 & 0.1268 & 0.0149 & 7.9921 & 0.9439 & 0.0058 & 0.0038 \\
2M12053972+6255594 & 181.4155121 & 62.9331703 & 0.1504 & 0.0328 & 6.9261 & 1.3306 & 0.0043 & 0.0027 \\
2M10262280+5424269 & 156.5950165 & 54.4074936 & 0.1289 & 0.0291 & 8.1584 & 1.8269 & 0.0035 & 0.0023 \\
2M09451473+2328285 & 146.3113861 & 23.4746113 & 0.1630 & 0.0267 & 6.2914 & 0.9699 & 0.0065 & 0.0041 \\
\hline
\end{tabular}
\end{center}
\end{table*}

The APOGEE survey has generated high resolution near-infrared spectra for over
$10^5$ stars. These data provide valuable information on the studies for
Galactic Archaeology. In order to fully exploit these data sets in
six-dimension, we used a Bayesian method to derive the distances and
extinctions of individual stars by considering photometric and spectroscopic
information, as well as prior knowledge on the Milky Way.

All derived distances and extinctions will be made available as online data.
Table \ref{table:catalogue} shows the online data format.  The first column is
APOGEE ID, the fourth and fifth columns are derived distance and its associated
error respectively, while the eighth and ninth columns are estimated extinction
and its error respectively. The sixth and seventh columns are another distance
indicator, parallax and its error output by the Bayesian method respectively.

To assess the derived distances, we compared our Bayesian distances to those
from four independent measurements, the {\it Hipparcos} parallaxes, stellar
cluster distances, Red Clump star distances, asteroseismic distances from SAGA
catalogue \citep{casagrande14} and APOKASC  catalogue \citep{rodrigues14}.
These independent measurements cover four orders of magnitude in distance, from
$\sim 0.02$ kpc (the {\it Hipparcos} parallaxes) to $\sim 20$ kpc (star
clusters). The results of these validations are all summarized in Table
\ref{table:d}.  We find that the mean relative difference between our Bayesian
distances and those derived from other methods are $-4.2\%$ to $+3.6\%$, and
that the dispersion ranges from 15\% to 25\%.  Surprisingly, no systematic
offset is seen in the comparison with one set of asteroseismic constrained
distances \citep{casagrande14}. Considering mutual uncertainties in our derived
distances and in other measurements, we conclude that, statistically, our
derived distances are accurate to a few percent although errors for a few
individual stars could be large. Therefore, they are suitable for statistical
studies on Galactic Archaeology.

We have compared our Bayesian extinctions with those derived from the RJCE
method. It seems that the extinctions from the RJCE method depend on spectral
type. For stars with $\Teff < 4000$ K, $\logg < 1.0$, and $\mh < -1.0$, the
RJCE method tends to overestimate extinction, however, in most cases this
overestimation is less than 0.11 Mag. in $K_s-$band, which is still within the
error budget suggested by \citet{majewski11}. The overestimation of the
extinction at large distance can also be attributed to this effect, as giants
with even lower $\logg$ are usually the dominant population at this region due
to a selection effect. We also notice that there is a systematic difference at
heavy extinction $A(K_s)>0.8$ Mag., which could be due to the adopted
extinction law. Comparing with two dimensional extinction map of
\citet{gonzalez12}, we note that the map may overestimate extinction in the
nearby regions.

We also compared our results with three-dimensional extinction models
of \citet{drimmel03} and maps of \citet{marshall06}, \citet{schultheis14b}, and
\citet{chen14}. Compared with our extinction, both the model of
\citet{drimmel03} and map of \citet{chen14} show lower extinction values for
distant stars toward the anti-galactic central region. The model of
\citet{drimmel03} matches well with our results of nearby region $(d<7$kpc$)$
toward the bulge field $(l=0.5,b=-2.0)$. For this field the map of
\citet{marshall06} predicts a steep rise in slope for $d<4$ kpc, while the map
of \citet{schultheis14b} provides a good fit with our result for distant region
($d>7$ kpc).

The accurate distance and extinction derived in the current work provide
valuable information for the APOGEE stars, and are important for
exploiting the data set of the APOGEE survey when studying the formation
and evolution of the Milky Way.

\section*{Acknowledgments}

The anonymous referee is greatly thanked for useful comments and suggestions
that helped improve the quality of this paper.  We warmly thank Diane Feuillet
for sharing her {\it Hipparcos} target list before publishing it and Jon Holtzman for
the useful discussion. Haibo Yuan is thanked for useful discussion. This
research was supported by the National Key Basic Research Program of China
2014CB845700, and by the National Natural Science Foundation of China under
grant Nos. 11321064, 11390371, 11473033, 11428308 and U1331122.

This publication makes use of data products from the Two Micron All Sky Survey,
which is a joint project of the University of Massachusetts and the Infrared
Processing and Analysis Center/California Institute of Technology, funded by
the National Aeronautics and Space Administration and the National Science
Foundation.

This publication makes use of data products from the Wide-field Infrared Survey
Explorer, which is a joint project of the University of California, Los
Angeles, and the Jet Propulsion Laboratory/California Institute of Technology,
funded the National Aeronautics and Space Administration.

Funding for SDSS-III has been provided by the Alfred P. Sloan
Foundation, the Participating Institutions, the National Science
Foundation, and the US Department of Energy Office of Science.
The SDSS-III web site is http://www.sdss3.org/.

SDSS-III is managed by the Astrophysical Research Consor-
tium for the Participating Institutions of the SDSS-III Collabo-
ration including the University of Arizona, the Brazilian Partici-
pation Group, Brookhaven National Laboratory, Carnegie Mellon
University, University of Florida, the French Participation Group,
the German Participation Group, Harvard University, the Instituto
de Astrofisica de Canarias, the Michigan State/Notre Dame/JINA
Participation Group, Johns Hopkins University, Lawrence Berke-
ley National Laboratory, Max Planck Institute for Astrophysics,
Max Planck Institute for Extraterrestrial Physics, New Mexico State
University, New York University, Ohio State University, Pennsyl-
vania State University, University of Portsmouth, Princeton Uni-
versity, the Spanish Participation Group, University of Tokyo,
University of Utah, Vanderbilt University, University of Virginia,
University of Washington, and Yale University.

{}
\label{lastpage}
\end{document}